\documentclass[pdflatex,sn-basic]{sn-jnl} 

\usepackage{graphicx}%
\usepackage{multirow}%
\usepackage{amsmath,amssymb,amsfonts}%
\usepackage{amsthm}%
\usepackage{mathrsfs}%
\usepackage[title]{appendix}%
\usepackage{xcolor}%
\usepackage{textcomp}%
\usepackage{manyfoot}%
\usepackage{booktabs}%
\usepackage{algorithm}%
\usepackage{algorithmicx}%
\usepackage{algpseudocode}%
\usepackage{listings}%

\begin{document}

\title[Power spectral analysis of neurons]{Power spectral analysis of voltage-gated channels in neurons}

\author{\fnm{Christophe} \sur{Magnani}}
\author{\fnm{Lee E.} \sur{Moore}}
\affil{\\ \small{Centre Borelli, CNRS UMR 9010, Universit\'{e} Paris Cit\'{e}, France}}

\abstract{
This article develops a fundamental insight into the behavior of neuronal membranes, focusing on their responses to stimuli measured with power spectra in the frequency domain. It explores the use of linear and nonlinear (quadratic sinusoidal analysis) approaches to characterize neuronal function. It further delves into the random theory of internal noise of biological neurons and the use of stochastic Markov models to investigate these fluctuations. The text also discusses the origin of conductance noise and compares different power spectra for interpreting this noise. Importantly, it introduces a novel sequential chemical state model, named $p_{2}$, which is more general than the Hodgkin-Huxley formulation, so that the probability for an ion channel to be open does not imply exponentiation. In particular, it is demonstrated that the $p_{2}$ (without exponentiation) and $n^{4}$ (with exponentiation) models can produce similar neuronal responses. A striking relationship is also shown between fluctuation and quadratic power spectra, suggesting that voltage-dependent random mechanisms can have a significant impact on deterministic nonlinear responses, themselves known to have a crucial role in the generation of action potentials in biological neural networks.
}

\keywords{Hodgkin-Huxley, Markov, voltage-gated ion channels, neuronal noise, admittance, quadratic sinusoidal analysis}

\maketitle

\section{Introduction}\label{sec_introduction}

The purpose of this article is to elucidate the fundamental behavior of excitable neuronal membranes by using recent methods in the frequency domain. Since the historical work of the French mathematician and physicist \cite{Fourier1822} in The Analytical Theory of Heat, it is known that many kinds of signals admit a dual representation either as a real valued function $u\left(t\right)$ of the time variable $t$ or as a complex valued function $\widehat{u}\left(\omega\right)$ of the frequency variable $\omega$ where $\widehat{u}$ is the Fourier transform of $u$. Using this approach, individual excitable cells usually studied by their responses to constant stimuli can also be investigated by their responses to multi-sinusoidal stimuli over a broad frequency range.

With the current clamp technique, the reactions of individual cells to current stimuli generally show threshold impulses of the membrane potential that provide the means for signal transmission throughout the nervous system. Precisely because of this threshold property, it is difficult to determine the ionic currents underlying those action potential impulses. A major technical advance in measuring neuronal properties occurred with the advent of the voltage clamp technique \citep{Bear2016}, which was first invented by Cole and \cite{Marmont_1949} and later exploited by \cite{Hodgkin1952} who developed the Hodgkin-Huxley (HH) equations that have become the gold standard for most neuronal models in real time simulations. With the voltage clamp, a retroactive electronic device controls the membrane potential such that the neuronal properties can be quantitatively determined from the measured current elicited by a change in the potential.

Both current clamp and voltage clamp techniques can be used in the frequency domain to characterize neuronal function. A typical approach considers linear analysis by calculating impedance and admittance, as described by \cite{Mauro1970} for the squid giant axon. However, neuronal behavior is fundamentally nonlinear due to the voltage dependence of most ionic channels (for instance potassium or sodium). Quadratic sinusoidal analysis (QSA) is a recent method developed by \cite{Magnani2011} that provides a fundamental insight of the linear and quadratic neuronal behaviors using matrix calculus in the frequency domain. Concrete applications have been done with neurons involved in the oculo-vestibular integrator \citep{Magnani2013} as well as with neurons of the medial entorhinal cortex which are part of the grid cell network \citep{Magnani2014}. These linear and nonlinear approaches in the frequency domain are much more efficient and concise than time domain methods for extracting stationary and dynamic features from neurons by slightly perturbing them with multi-sinusoidal signals around a steady state.

Although such a smooth deterministic description by frequency waves is able to capture fundamental properties of the neuronal function, biological neurons are significantly perturbed by internal noise. Among the different kinds of noise sources described by \cite{Stevens_1972}, conductance fluctuations reflect the stochastic nature of ionic channels at the microscopic level. For this reason, stochastic Markov models have been used in this article to investigate the intrinsic fluctuations and their relationships to the complicated nonlinear behavior of neurons. This approach extends QSA analysis with stochastic Markov simulations and compares the power spectra of linear, quadratic and stochastic neuronal processes.

Some of the earliest measurements of membrane voltage noise were done on the node of Ranvier by \cite{Verveen1968} who suggested that the opening and closing of ionic channels lead to voltage fluctuations. Later, voltage clamp measurements on squid axons suggested that current channel noise was filtered by the axonal membrane that can be measured as a voltage power spectrum. Extensive measurements have been made of spontaneous fluctuations under both current and voltage clamp conditions along with linear impedance analysis to assess the basis of the measured noise power spectrum \citep{Fishman_1983,Conti1975,Conti1984,Poussart1969,Poussart1977}.

An early and perceptive paper by \cite{Stevens_1972} showed that the Hodgkin-Huxley equations themselves provide two different interpretations of the origin of conductance noise, namely the opening and closing of whole conductance channel whose opening is controlled by multiple gating particles ($n$, $m$ or $h$), or alternatively, fluctuations of individual gating particles. The first case involves probability and correlation functions having multiple terms. For the potassium conductance where $g_{\mathrm{K}}=g_{\mathrm{K}}^{\mathrm{max}}n^{4}$, this leads to conductance noise power spectra consisting of four Lorentzian terms. In the second case, the Hodgkin-Huxley equations are linearized and the potassium conductance power spectrum consists of a single term related to the potassium channel time constant. In the first case, the nonlinear properties of the channel ($n^{4}$) would be involved in the origin of the spontaneous fluctuation, while in the second case, the linearized impedance ($n$) would likely be a good predictor of the voltage noise.

The measured squid axon spontaneous noise and the nonlinear power spectra of the Hodgkin-Huxley model both show resonance in the voltage measurements and non-resonating Lorentzian functions under voltage clamp. There is strong experimental and theoretical evidence that the spontaneous conductance noise cannot be predicted by the linear response from the same axon \citep{Fishman_1983,Poussart1969}.

With the advent of measurements of single ionic channels in neuronal membranes, many of the detailed properties of different ionic channels as well as their macromolecular basis have dominated excitable membrane research. These findings have stimulated the development of stochastic ionic conductance models, again using the Hodgkin-Huxley equations as a fundamental basis. The gold standard method to simulate the stochastic behavior appears to be Markov models which provide fluctuation noise power spectra identical to the first Hodgkin-Huxley interpretation by Stevens discussed above. Numerous papers have derived stochastic fluctuation power spectra based on this nonlinear character \citep[for instance]{ODonnell2014,Goldwyn2011b}. In addition to squid axon, measurements from nodes of Ranvier \citep{Conti1975,Conti1984,Sigworth1980,Elinder2001} and other preparations are consistent with the nonlinear origin of ionic power spectra. The interpretation, namely that certain nonlinear properties can show a probabilistic or stochastic character, suggests that they are involved in the fundamental origins of spontaneous channel noise in neurons. Since this clearly indicates that the stochastic behavior of excitable membranes is nonlinear, it is useful to quantitatively compare the power spectrum content of the nonlinear QSA responses with the corresponding simulations of Markov models. This will be done rigorously from the Hodgkin-Huxley equations described in the methods.

In the methods section, the Hodgkin-Huxley model is briefly reviewed. Linear analysis in the frequency domain is introduced based on the work of \cite{Mauro1970}. The quadratic sinusoidal analysis (QSA) is introduced in two ways, first the basic theory as described by \cite{Magnani2011} and second, a new algorithm of frequency averaging to calculate power spectra. Fluctuation simulations are done with Markov models based on the work of \cite{Goldwyn2011a,Goldwyn2011b}. The theoretical expressions for fluctuation power spectra are derived from the work of \cite{ODonnell2014}.

In the results section, the exponentiation of the potassium $n^{4}$ model is reduced to the minimal degree of nonlinearity $n^{2}$. Then, a novel sequential chemical state model named $p_{2}$ is introduced as a generalization of the $n^{2}$ model without exponentiation. Linear analysis, fluctuation simulations and theoretical power spectra are applied to the $p_{2}$ model by adapting the previous methods. Linear and quadratic functions are compared to neuronal fluctuations by adapting the previous methods. The $p_{2}$ and $n^{4}$ models are compared and it is demonstrated that they can produce similar neuronal responses. Remarkably, it is illustrated how the fluctuation-dissipation theorem can be violated at depolarized membrane potentials. The time constants of the $p_{2}$ and $n^{4}$ models seem more consistent with stochastic analysis than linear analysis.

Finally, the discussion section deals with the origin of fluctuations in the nonlinear neuronal responses. Surprisingly, simulations show that, for certain stimulus amplitudes and membrane surfaces, random voltage-dependent fluctuations can significantly modify deterministic nonlinear responses.

\section{Methods}\label{sec_methods}

\subsection{Hodgkin-Huxley model}\label{subsec_hodgkin_huxley_model}

The standard Hodgkin-Huxley model was originally proposed by \cite{Hodgkin1952} to describe the initiation and propagation of action potentials in the squid giant axon based on voltage clamp experiments. In this model, the current across the lipid bilayer is defined by
\[
C_{m}\frac{dV}{dt}=I-I_{\mathrm{L}}-I_{\mathrm{K}}-I_{\mathrm{Na}}
\]
where $C_{m}$ is the membrane capacitance, $V$ is the membrane potential, $I$ is the total membrane current, $I_{\mathrm{L}}$ is the leak current, $I_{\mathrm{K}}$ is the current through potassium ion channels and $I_{\mathrm{Na}}$ is the current through sodium ion channels.

The ionic currents are expressed with conductances
\begin{eqnarray*}
	I_{\mathrm{L}} & = & g_{\mathrm{L}}\left(V-V_{\mathrm{L}}\right)\\*
	I_{\mathrm{K}} & = & g_{\mathrm{K}}n^{4}\left(V-V_{\mathrm{K}}\right)\\*
	I_{\mathrm{Na}} & = & g_{\mathrm{Na}}m^{3}h\left(V-V_{\mathrm{Na}}\right)
\end{eqnarray*}
where $g_{\mathrm{L}}$ is the leak conductance, $g_{\mathrm{K}}$ is the maximum potassium conductance and $g_{\mathrm{Na}}$ is the maximum sodium conductance. The constants $V_{\mathrm{L}}$, $V_{\mathrm{K}}$ and $V_{\mathrm{Na}}$ are reversal potentials for leak, potassium ion channel and sodium ion channel respectively. The gating variables $n$, $m$ and $h$ represent potassium channel activation, sodium channel activation and sodium channel inactivation respectively, their values are constrained between $0$ and $1$.

The kinetics of gating variables satisfies the following first order differential equations determined by pairs of rate constants $\alpha_{i}\left(V\right)$ and $\beta_{i}\left(V\right)$
\begin{eqnarray*}
	\frac{dn}{dt} & = & \alpha_{n}\left(V\right)\left(1-n\right)-\beta_{n}\left(V\right)n\\*
	\frac{dm}{dt} & = & \alpha_{m}\left(V\right)\left(1-m\right)-\beta_{m}\left(V\right)m\\*
	\frac{dh}{dt} & = & \alpha_{h}\left(V\right)\left(1-h\right)-\beta_{h}\left(V\right)h
\end{eqnarray*}

At the steady state corresponding to a membrane potential level $V_{0}$, the time derivatives vanish with constants
\begin{eqnarray*}
	n_{\infty}\left(V_{0}\right) & = & \alpha_{n}\left(V_{0}\right)\tau_{n}\left(V_{0}\right)\\*
	m_{\infty}\left(V_{0}\right) & = & \alpha_{m}\left(V_{0}\right)\tau_{m}\left(V_{0}\right)\\*
	h_{\infty}\left(V_{0}\right) & = & \alpha_{h}\left(V_{0}\right)\tau_{h}\left(V_{0}\right)
\end{eqnarray*}
where
\[
\tau_{n}=\left(\alpha_{n}+\beta_{n}\right)^{-1},\;\tau_{m}=\left(\alpha_{m}+\beta_{m}\right)^{-1},\;\tau_{h}=\left(\alpha_{h}+\beta_{h}\right)^{-1}
\]

\subsection{Linear analysis}\label{subsec_linear_analysis}

\cite{Cole1941} was among the first to use the frequency domain to investigate neurons and suggested that the linear response of the axon membrane could be modeled with equivalent circuit elements such as inductances, capacitances and resistances. Hodgkin and Huxley also showed that their findings were consistent with the potassium conductance being described as an inductive reactance. Later, a mathematical equivalence between nonlinear conductance Hodgkin-Huxley models and electrical circuits was developed by \cite{Mauro1970} and experimentally confirmed on the squid axon.

In papers published by \cite{Mauro1970} and \cite{Fishman1977}, the linearization of the Hodgkin-Huxley equations in the frequency domain is obtained for small perturbations at steady state. More precisely, let $X=\left(V,n,m,h\right)$ be a dynamic state, $X_{0}=\left(V_{0},n_{0},m_{0},h_{0}\right)$ a steady state and $\frac{dX}{dt}=F\left(X\right)$ the system of differential equations. Clearly, $n_{0}=n_{\infty}\left(V_{0}\right)$, $m_{0}=m_{\infty}\left(V_{0}\right)$ and $h_{0}=h_{\infty}\left(V_{0}\right)$ where $V_{0}$ can be controlled directly in voltage clamp or indirectly in current clamp. Then, differential calculus linearizes the system for a small perturbation steady state
\[
\delta F=F\left(X\right)-F\left(X_{0}\right)\sim\delta X^{T}DF\left(X_{0}\right)
\]

It follows linearization of the kinetic equations, for example for the potassium variable
\[
\delta\frac{dn}{dt}=\frac{d\alpha_{n}}{dV}\delta V-\alpha_{n}\delta n-n_{0}\frac{d\alpha_{n}}{dV}\delta V-\beta_{n}\delta n-n_{0}\frac{d\beta_{n}}{dV}\delta V
\]

This leads to the first order linear differential equation
\[
\frac{d\delta n}{dt}+\left(\alpha_{n}+\beta_{n}\right)\delta n=\left[\frac{d\alpha_{n}}{dV}-n_{0}\frac{d\left(\alpha_{n}+\beta_{n}\right)}{dV}\right]\delta V
\]

Applying the Fourier transform
\[
i\omega\widehat{\delta n}+\left(\alpha_{n}+\beta_{n}\right)\widehat{\delta n}=\left[\frac{d\alpha_{n}}{dV}-n_{0}\frac{d\left(\alpha_{n}+\beta_{n}\right)}{dV}\right]\widehat{\delta V}
\]

This can be solved in the frequency domain as
\[
\widehat{\delta n}=\frac{\frac{d\alpha_{n}}{dV}-n_{0}\frac{d\left(\alpha_{n}+\beta_{n}\right)}{dV}}{i\omega+\alpha_{n}+\beta_{n}}\widehat{\delta V}
\]
and similarly for $\widehat{\delta m}$ and $\widehat{\delta h}$. In order to simplify notations, this fraction will be denoted as $\widehat{D_{n}}$, $\widehat{D_{m}}$ and $\widehat{D_{h}}$ respectively.

Similarly, linearization of the ionic currents is given by
\begin{eqnarray*}
	\delta I_{\mathrm{L}} & = & g_{\mathrm{L}}\delta V\\*
	\delta I_{\mathrm{K}} & = & 4g_{\mathrm{K}}n_{0}^{3}\left(V_{0}-V_{\mathrm{K}}\right)\delta n+g_{\mathrm{K}}n_{0}^{4}\delta V\\*
	\delta I_{\mathrm{Na}} & = & 3g_{\mathrm{Na}}m_{0}^{2}h_{0}\left(V_{0}-V_{\mathrm{Na}}\right)\delta m+g_{\mathrm{Na}}m_{0}^{3}\left(V_{0}-V_{\mathrm{Na}}\right)\delta h+g_{\mathrm{Na}}m_{0}^{3}h_{0}\delta V
\end{eqnarray*}
which can be solved in the frequency domain as
\begin{eqnarray*}
	\widehat{\delta I_{\mathrm{L}}} & = & g_{\mathrm{L}}\widehat{\delta V}\\*
	\widehat{\delta I_{\mathrm{K}}} & = & g_{\mathrm{K}}\left[4n_{0}^{3}\left(V_{0}-V_{\mathrm{K}}\right)\widehat{D_{n}}+n_{0}^{4}\right]\widehat{\delta V}\\*
	\widehat{\delta I_{\mathrm{Na}}} & = & g_{\mathrm{Na}}\left[3m_{0}^{2}h_{0}\left(V_{0}-V_{\mathrm{Na}}\right)\widehat{D_{m}}+m_{0}^{3}\left(V_{0}-V_{\mathrm{Na}}\right)\widehat{D_{h}}+m_{0}^{3}h_{0}\right]\widehat{\delta V}
\end{eqnarray*}

Similarly, linearization of the current across the lipid bilayer is given by
\[
C_{m}\frac{d\delta V}{dt}=\delta I-\delta I_{\mathrm{L}}-\delta I_{\mathrm{K}}-\delta I_{\mathrm{Na}}
\]
or equivalently in the frequency domain
\[
C_{m}i\omega\widehat{\delta V}=\widehat{\delta I}-\widehat{\delta I_{\mathrm{L}}}-\widehat{\delta I_{\mathrm{K}}}-\widehat{\delta I_{\mathrm{Na}}}
\]

The linearized response current $\widehat{\delta I}$ to the linearized stimulus voltage $\widehat{\delta V}$ is characterized by the admittance
\begin{equation}
	\widehat{Y}=\frac{\widehat{\delta I}}{\widehat{\delta V}}\label{eq:Y-definition}
\end{equation}
\begin{eqnarray}
	\widehat{Y} & = & C_{m}i\omega\label{eq:mauro_hh_admittance}\\*
	& + & g_{\mathrm{L}}\nonumber \\*
	& + & g_{\mathrm{K}}\left[4n_{0}^{3}\left(V_{0}-V_{\mathrm{K}}\right)\widehat{D_{n}}+n_{0}^{4}\right]\nonumber \\*
	& + & g_{\mathrm{Na}}\left[3m_{0}^{2}h_{0}\left(V_{0}-V_{\mathrm{Na}}\right)\widehat{D_{m}}+m_{0}^{3}\left(V_{0}-V_{\mathrm{Na}}\right)\widehat{D_{h}}+m_{0}^{3}h_{0}\right]\nonumber 
\end{eqnarray}

Equivalently, the impedance is defined as
\begin{equation}
	\widehat{Z}=\frac{1}{\widehat{Y}}\label{eq:mauro_hh_impedance}
\end{equation}

Note that the frequency $f$ in Hz is interchangeable with angular frequency $\omega=2\pi f$, both are used in this article depending on the context.

\begin{figure}[H]
\centering
\includegraphics[width=12cm]{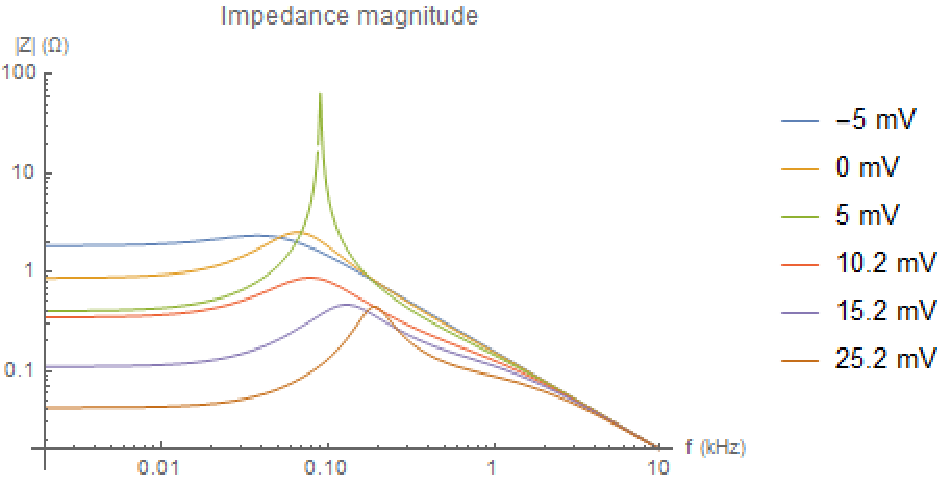}
\caption{\label{fig:linear_k_na_impedance}\textbf{Impedance functions for the Hodgkin-Huxley model.} Different membrane potential displacements $V_{0}$ were applied from the resting potential of zero. Curves from top to bottom at lowest frequencies: $-5$, $0$, $5$, $10.2$, $15.2$, $25.2$ in mV. Abscissa: frequency $f$ in Hz (logarithmic scale). Ordinate: magnitude $\left|\widehat{Z}\left(f,V_{0}\right)\right|$ in $\Omega$ (logarithmic scale). Parameters of the simulation are given in Table \ref{tab_linear_k_na_parameters}.}
\end{figure}

One of the most striking features of neurons and their models is the presence of an impedance resonance at certain membrane potentials $V_{0}$. This is illustrated for the voltage clamped Hodgkin-Huxley model especially by the two amplitude peaks at $5$ mV and $25.2$ mV depolarizations shown in Fig. \ref{fig:linear_k_na_impedance}. The plots are superimposed magnitudes $\left|\widehat{Z}\left(f,V_{0}\right)\right|$ of the Eq. \ref{eq:mauro_hh_impedance}, which is identical to a small signal sinusoidal stimulus of the full nonlinear Hodgkin-Huxley equations. The simulation results are similar to actual measurements on squid axons independent of the electrode properties.

\begin{table}[h]
\caption{Parameters of the simulation of the Hodgkin-Huxley equations used to compute impedances given by Eq. \ref{eq:mauro_hh_impedance} at different membrane potential displacements $V_{0}$.}\label{tab_linear_k_na_parameters}
\begin{tabular}{|c|p{8cm}|}
\hline
$C_{m}$ & $1\textrm{\ensuremath{\mu}\ensuremath{\mathrm{F}}}/\mathrm{cm}^{2}$\tabularnewline
\hline
$g_{\mathrm{L}}$ & $0.3\mathrm{mS}/\mathrm{cm}^{2}$\tabularnewline
\hline
$V_{\mathrm{L}}$ & $10.6\mathrm{mV}$\tabularnewline
\hline
$g_{\mathrm{K}}$ & $36\mathrm{mS}/\mathrm{cm}^{2}$\tabularnewline
\hline
$V_{\mathrm{K}}$ & $-12\mathrm{mV}$\tabularnewline
\hline
$g_{\mathrm{Na}}$ & $120\mathrm{mS}/\mathrm{cm}^{2}$\tabularnewline
\hline
$V_{\mathrm{Na}}$ & $120\mathrm{mV}$\tabularnewline
\hline
$\alpha_{n}\left(V\right)$ & $0.01\cdot\left(10-V\right)/\left(e^{\left(10.001-V\right)/10}-1\right)$\tabularnewline
\hline
$\beta_{n}\left(V\right)$ & $0.125\cdot e^{-V/80}$\tabularnewline
\hline
$\alpha_{m}\left(V\right)$ & $0.1\cdot\left(25-V\right)/\left(e^{\left(25-V\right)/10}-1\right)$\tabularnewline
\hline
$\beta_{m}\left(V\right)$ & $4\cdot e^{-V/18}$\tabularnewline
\hline
$\alpha_{h}\left(V\right)$ & $0.07\cdot e^{-V/20}$\tabularnewline
\hline
$\beta_{h}\left(V\right)$ & $1/\left(e^{\left(30-V\right)/10}+1\right)$\tabularnewline
\hline
Stimulus amplitude & $0.0125\mathrm{mV}$ for Hodgkin-Huxley model, default $0.25\mathrm{mV}$ for $n^{4}$ and $p_{2}$ models\tabularnewline
\hline
Stimulus frequencies & $0.2$, $0.7$, $2$, $3$, $10$, $21$, $35$, $50$, $76$, $104$,
$134$, $143$, $223$, $239$, $285$, $388$, $405$, $515$, $564$,
$636$, $815$, $892$, $982$ (units in $\mathrm{Hz}$)\tabularnewline
\hline
Stimulus phases & Pseudo-random values $\left[0;\pi\right]$ $\mathrm{rad}$\tabularnewline
\hline
\end{tabular}
\end{table}

Although impedance resonance is a linear property of the Hodgkin-Huxley equations, it is due to the voltage dependence of the steady state values of the ionic conductances. In particular, $\frac{dn_{\text{\ensuremath{\infty}}}}{dv}>0$ for the potassium conductance (i.e. $n_{\infty}\left(V_{0}\right)$ does increase with depolarizing levels $V_{0}$). In Eq. \ref{eq:mauro_hh_admittance}, the potassium term in $g_{\mathrm{K}}$ was shown to be equivalent to an inductive reactance by \cite{Mauro1970}. Similarly the sodium term $g_{\mathrm{Na}}$ can be described by other circuit elements. Thus, the Hodgkin-Huxley model or any excitable cell can be analyzed by a piecewise linear analysis at different membrane potentials as shown in Fig. \ref{fig:linear_k_na_impedance}. Data collected in this manner over a range of membrane potentials allows one to determine the voltage dependence of the active conductances in addition to passive properties, and in turn construct a system of nonlinear differential equations for a particular model. Thus, a further advantage of frequency domain measurements is that model discrimination using parameter estimation can be more accurate if both real time and impedance results are used, as shown by \cite{Murphey1995}.

\begin{figure}[H]
\centering
\includegraphics[width=12cm]{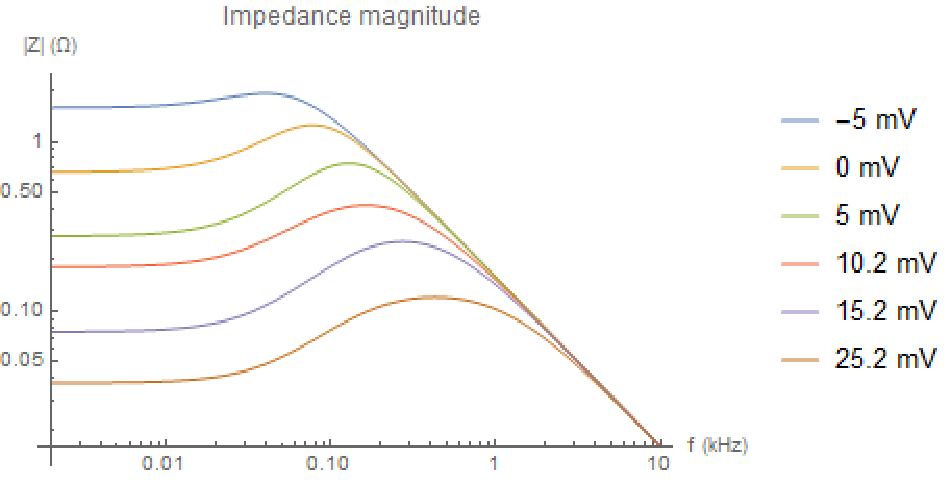}
\caption{\label{fig:linear_k_impedance}\textbf{Impedance functions for the Hodgkin-Huxley model when $g_{\mathrm{Na}}=0$.} Different membrane potential displacements $V_{0}$ were applied from the resting potential of zero. Curves from top to bottom at lowest frequencies: $-5$, $0$, $5$, $10.2$, $15.2$, $25.2$ in mV. Abscissa: frequency $f$ in Hz (logarithmic scale). Ordinate: magnitude $\left|\widehat{Z}\left(f,V_{0}\right)\right|$ in $\Omega$ (logarithmic scale). Parameters of the simulation are given in Table \ref{tab_linear_k_na_parameters} with $g_{\mathrm{Na}}=0$.}
\end{figure}

Since neurons are composed of a minimum of two conductances, inward sodium or calcium currents and various outward potassium and other currents, it is useful to consider their individual contributions to the frequency domain behavior. This paper is focused on the potassium conductance as a model of any of the individual conductances, thus $g_{\mathrm{Na}}=0$. Fig. \ref{fig:linear_k_impedance} illustrates one broad impedance resonance maximum for the Hodgkin-Huxley potassium conductance alone, which shifts to higher frequencies with increasing depolarizations. In contrast, the impedances of Fig. \ref{fig:linear_k_na_impedance} have sharper resonances and two peaks ($5$ mV and $25.2$ mV) clearly due to the presence of the sodium conductance in conjunction with potassium. Thus, one useful aspect of this analysis is that the number of resonance peaks can give an indication of the minimum number of active conductance processes present.

\subsection{Quadratric analysis}\label{subsec_quadratic_analysis}

Biological neurons and their models are fundamentally nonlinear systems that sometimes significantly contradict the linear superposition principle. System identification methods such as Volterra and Wiener series have been widely used to characterize nonlinear neuronal functions. In the pioneering work of \cite{Marmarelis1972}, Wiener theory was applied to predict the nonlinear behavior of a neuron chain in the catfish retina. Unfortunately, it is practically difficult and time-consuming to calculate the kernel coefficients of the Volterra and Wiener series, as these methods generally require extensive data analysis, analogous to averaging a small signal from extraneous noise.

In general, the use of random broad band stimuli similar to a white noise is time consuming, even for linear methods requiring averaging over experiments. However, significant reduction of experimental time is possible if a pseudo random stimulus containing a limited number of frequencies is applied to the preparation \citep{Fishman1977,Poussart1969}. Responses to such stimuli have an excellent signal to noise ratio and little or no averaging is necessary when measuring the impedance of a single cell.

In the case of nonlinear systems, multi-sinusoidal stimuli can be used to precisely measure deterministic linear and quadratic responses, provided that generated output frequencies do not overlap at first and second orders. For example, input frequencies $f_{1}=1$, $f_{2}=2$, $f_{3}=3$, $f_{4}=4$ (in Hz) generate unwanted overlaps because $f_{1}+f_{2}=f_{3}$ or $f_{1}+f_{4}=f_{2}+f_{3}$. Multi-sinusoidal stimuli with selected frequencies to avoid overlap at first and second orders were used by \cite{Magnani2011} to study subthreshold neuronal responses, as well as previously by \cite{Victor1980} to study cat retinal ganglion cells.

Such non-overlapping multi-sinusoidal stimuli must have sufficiently large amplitude to elicit both linear and quadratic responses, while remaining sufficiently small to avoid higher order contamination. The quadratic sinusoidal analysis, termed QSA, was introduced by \cite{Magnani2011} to provide a flexible way to capture the linear and quadratic neuronal functions at subthreshold membrane potentials, as well as to compare biological experiments with theoretical models \citep{Magnani2013,Magnani2014}. In particular, the subthreshold membrane potential just below the threshold is fundamentally a nonlinear process which is critically involved in action potential generation.

It is well known that a sinusoidal signal can be expressed as a sum of complex exponentials in Fourier analysis, for example :
\[
\cos\left(\omega t\right)=\frac{1}{2}e^{i\omega t}+\frac{1}{2}e^{-i\omega t}
\]
where $i^{2}=-1$ and $\omega$ is the angular frequency. More generally, a multi-sinusoidal signal of $N$ frequencies can be expressed as a sum of complex exponentials
\[
x\left(t\right)=\sum_{k\in\Gamma}x_{k}e^{i\omega_{k}t}
\]
where $\Gamma=\left\{ -N,\cdots,-1,+1,\cdots,+N\right\} $ enumerates integers between $-N$ and $+N$ (zero excluded), $x_{k}$ are complex Fourier coefficients such that $x_{-k}=\overline{x_{k}}$ and $\omega_{k}$ are angular frequencies such that $\omega_{-k}=-\omega_{k}$. The duration $T$ (seconds) of the experiment determines the lowest frequency $\frac{1}{T}$ (Hertz). The angular frequencies are defined by integer multiples of the lowest frequency, that is to say $\omega_{k}=2\pi\frac{n_{k}}{T}$ where $n_{k}$ are integers such that $n_{-k}=-n_{k}$. Each Fourier coefficient is a complex number that can be decomposed as $x_{k}=\left|x_{k}\right|e^{i\theta_{k}}$ where $\left|x_{k}\right|$ is the amplitude (nonnegative real number) and $\theta_{k}$ is the phase (between $0$ and $2\pi$). It is good practice to randomize the phases of multi-sinusoidal stimuli to avoid biases in neuronal responses.

Such a multi-sinusoidal stimulus $x\left(t\right)$ can be applied to neuronal cells in voltage clamp and current clamp experiments. Assuming that quality criteria for the QSA method are satisfied \citep{Magnani2011}, the output signal measured by the electrode can be decomposed as
\[
y\left(t\right)=y_{0}+y_{1}\left(t\right)+y_{2}\left(t\right)
\]
where $y_{0}$ is the DC component, $y_{1}\left(t\right)$ is the linear component and $y_{2}\left(t\right)$ is the quadratic component. More specifically
\begin{equation}
	y\left(t\right)=y_{0}+\sum_{k\in\Gamma}l_{k}x_{k}e^{i\omega_{k}t}+\sum_{i,j\in\Gamma}b_{i,j}x_{i}e^{i\omega_{i}t}x_{j}e^{i\omega_{j}t}\label{eq:qsa_analytic}
\end{equation}
where $l_{k}$ and $b_{i,j}$ are complex numbers characterizing the neuronal response to the stimulus, by similarity with impedance or admittance. By convention, $b_{-k,k}=0$ for all $k\in\Gamma$ so that all DC components are encoded in $y_{0}$.

The QSA theory is based on a vector representation of multi-sinusoidal signals in an orthonormal vector basis $\left\{ \mathbf{e_{k}}\right\} _{k\in\Gamma}$ representing the complex exponentials $\left\{ e^{i\omega_{k}t}\right\} _{k\in\Gamma}$. In this way, the multi-sinusoidal stimulus $x\left(t\right)$ is encoded as a time independent vector
\[
\mathbf{x}=\sum_{k\in\Gamma}x_{k}\mathbf{e_{k}}
\]

The corresponding time dependent vectors are defined by
\[
\mathbf{x}_{t}=\sum_{k\in\Gamma}x_{k}e^{i\omega_{k}t}\mathbf{e_{k}}
\]

Putting the coefficients $l_{k}$ in a row matrix $\mathbf{L}$ and the coefficients $b_{i,j}$ in a square matrix $\mathbf{B}$, Eq. \ref{eq:qsa_analytic} can be reformulated with linear algebra as
\begin{equation}
	y\left(t\right)=y_{0}+\mathbf{L}\mathbf{x}_{t}+\mathbf{x}_{t}^{T}\mathbf{B}\mathbf{x}_{t}\label{eq:qsa_algebraic}
\end{equation}

By noticing that
\begin{eqnarray*}
	\sum_{i,j\in\Gamma}b_{i,j}x_{i}e^{i\omega_{i}t}x_{j}e^{i\omega_{j}t} & = & \sum_{i,j\in\Gamma}b_{-i,j}\overline{x_{i}e^{i\omega_{i}t}}x_{j}e^{i\omega_{j}t}
\end{eqnarray*}
we are led to define the QSA matrix $\mathbf{Q}$ by
\[
Q_{i,j}=B_{-i,j}
\]
so that
\[
y\left(t\right)=y_{0}+\mathbf{L}\mathbf{x}_{t}+\overline{\mathbf{x}}_{t}^{T}\mathbf{Q}\mathbf{x}_{t}
\]

Remarkably, the QSA matrix $\mathbf{Q}$ is Hermitian \citep{Magnani2011}, which means that
\[
\overline{\mathbf{Q}}^{T}=\mathbf{Q}
\]

This algebraic approach has been widely used in modern physics, especially in quantum physics where Hermitian operators represent physical observables. In particular, the eigenvalues of the QSA matrix are real numbers which form a spectrum that can be interpreted as a signature of the quadratic neuronal function.

The most glaring property of the quadratic neuronal response is the generation of second order frequencies that are not present in the stimulus. More precisely, Eq. \ref{eq:qsa_analytic} shows that the linear components $l_{k}x_{k}e^{i\omega_{k}t}$ contain only the stimulus frequencies $\omega_{k}$, while the quadratic components $b_{i,j}x_{i}e^{i\omega_{i}t}x_{j}e^{i\omega_{j}t}=b_{i,j}x_{i}x_{j}e^{i\left(\omega_{i}+\omega_{j}\right)t}$ contain new frequencies $\omega_{i}+\omega_{j}$ that are not present in the stimulus in the absence of overlap. The row matrix $\mathbf{L}$ and the QSA matrix $\mathbf{Q}$ can be interpreted as linear and quadratic filters respectively, by extension of the impedance or admittance concept.

\subsection{Multi-sinusoidal power spectra}\label{subsec_multisinusoidal_power_spectra}

The concept of power spectrum is useful in statistical analysis, both for signal processing and for stochastic processes, so it is natural to use it in this paper. The power spectrum of a multi-sinusoidal signal $u\left(t\right)$ is given by $\left|\hat{u}\left(\omega\right)\right|^{2}$ where $\widehat{u}$ is the Fourier transform of $u$ and $\omega$ is the frequency variable. The power spectrum can be averaged over a set of measurements.

For a single measurement of the output $y\left(t\right)$ in response to a stimulus $x\left(t\right)$ with fundamental frequencies $\left|\omega_{k}\right|$, the multi-sinusoidal power spectra cover the positive frequencies of linear and quadratic analyses:
\begin{itemize}
\item Linear power spectrum at fundamental frequencies:
\[
S_{L}\left(\left|\omega_{k}\right|\right)=\left|\hat{y}_{1}\left(\left|\omega_{k}\right|\right)\right|^{2}
\]
\item Quadratic power spectrum at frequency doubling:
\[
S_{D}\left(2\left|\omega_{k}\right|\right)=\left|\hat{y}_{2}\left(2\left|\omega_{k}\right|\right)\right|^{2}
\]
\item Quadratic power spectrum at frequency sums for $\left|\omega_{i}\right|\neq\left|\omega_{j}\right|$:
\[
S_{P}\left(\left|\omega_{i}\right|+\left|\omega_{j}\right|\right)=\left|\hat{y}_{2}\left(\left|\omega_{i}\right|+\left|\omega_{j}\right|\right)\right|^{2}
\]
\item Quadratic power spectrum at frequency differences for $\left|\omega_{i}\right|\neq\left|\omega_{j}\right|$:
\[
S_{M}\left(\left|\left|\omega_{i}\right|-\left|\omega_{j}\right|\right|\right)=\left|\hat{y}_{2}\left(\left|\left|\omega_{i}\right|-\left|\omega_{j}\right|\right|\right)\right|^{2}
\]
\end{itemize}

The quadratic power spectra are indexed by second order frequencies, which are not stimulus frequencies because they were chosen without overlap. Thus, it would be convenient to have also a quadratic power spectrum indexed by fundamental frequencies as an alternative representation. To this end, a function similar to the ``R summation function'' of \cite{Magnani2013} is introduced in a different way below, computing the mean squared quadratic output by matrix columns:
\[
S_{R}\left(\omega_{j}\right)=\frac{1}{2N}\sum_{i\in\Gamma}\left|Q_{i,j}\overline{x_{i}}x_{j}\right|^{2}
\]

Although these multi-sinusoidal power spectra reflect exact neuronal responses, they are not accurate because they are defined over a small set of non-overlapping frequencies. Therefore, it is necessary to average multiple measurements to increase the accuracy of the power spectra.

To perform this averaging, a set of $M$ stimuli $x^{(m)}\left(t\right)$ is generated with sets of non-overlapping random frequencies $\Omega^{(m)}$ for $m=1,\ldots,M$. Each set of first order frequencies $\Omega^{(m)}$ determines a set of second order frequencies $\Xi^{(m)}$. The global sets of first and second order frequencies are defined by merging the individual sets, respectively
\begin{eqnarray*}
	\Omega & = & \bigcup_{m=1}^{M}\Omega^{(m)}\\*
	\Xi & = & \bigcup_{m=1}^{M}\Xi^{(m)}
\end{eqnarray*}

Importantly, different individual sets of the same order can share frequencies, so it is necessary to count redundancies
\begin{eqnarray*}
	\mathcal{N}_{\Omega}\left(\omega\right) & = & \#\left\{ m\vert\omega\in\Omega^{(m)}\right\} \\*
	\mathcal{N}_{\Xi}\left(\xi\right) & = & \#\left\{ m\vert\xi\in\Xi^{(m)}\right\} 
\end{eqnarray*}
where $\#$ indicates the number of elements in a set.

Fourier analysis of the measured linear responses $y_{1}^{(m)}\left(t\right)$ yields output sets $\hat{y}_{1}^{(m)}\left(\omega\right)$ defined at the first order frequencies $\omega\in\Omega^{(m)}$ and zero elsewhere. This allows to calculate the averaged linear power spectrum for $\omega\in\Omega$
\[
S_{L}\left(\omega\right)=\frac{1}{\mathcal{N}_{\Omega}\left(\omega\right)}\sum_{m=1}^{M}\left|\hat{y}_{1}^{(m)}\left(\omega\right)\right|^{2}
\]
where $\omega\in\Omega$ implies $\mathcal{N}_{\Omega}\left(\omega\right)\geq1$ so the denominator is not zero.

Fourier analysis of the measured quadratic responses $y_{2}^{(m)}\left(t\right)$ yields output sets $\hat{y}_{2}^{(m)}\left(\xi\right)$ defined at the second order frequencies $\xi\in\Xi^{(m)}$ and zero elsewhere. This allows to calculate the averaged quadratic power spectrum for $\xi\in\Xi$
\[
S_{2}\left(\xi\right)=\frac{1}{\mathcal{N}_{\Xi}\left(\xi\right)}\sum_{m=1}^{M}\left|\hat{y}_{2}^{(m)}\left(\xi\right)\right|^{2}
\]
where $\xi\in\Xi$ implies $\mathcal{N}\left(\xi\right)\geq1$ so the denominator is not zero. The partial quadratic power spectra $S_{D}$, $S_{P}$, $S_{M}$ can be calculated using the same method.

The power spectra $S_{R}$ are calculated on first order frequencies using the same method for $\omega\in\Omega$
\[
S_{R}\left(\omega\right)=\frac{1}{\mathcal{N}_{\Omega}\left(\omega\right)}\sum_{m=1}^{M}S_{R}^{(m)}\left(\omega\right)
\]
In particular, $\omega\in\Omega$ implies $\mathcal{N}_{\Omega}\left(\omega\right)\geq1$ so the denominator is not zero.

It should be noted that although different sets $\Omega^{(m)}$ and $\Xi^{(m)}$ may share frequencies, such post-result frequency redundancy is not frequency overlap, with each individual measurement $y^{(m)}\left(t\right)$ being the response to a stimulus without overlap.

\subsection{Stochastic Markov simulations}\label{subsec_stochastic_markov_simulations}

Although the Hodgkin-Huxley model is empirical, it has led to various biophysical interpretations. The potassium ion channel is generally considered to have four independent identical subunits, each characterized by a two-state process
\[
0\overset{\alpha}{\underset{\beta}{\rightleftharpoons}}1
\]
where $\alpha_{n}\left(V\right)$ and $\beta_{n}\left(V\right)$ are voltage-dependent transition rates given in the Hodgkin-Huxley model. All four subunits must be open for the channel to be open. If $n$ denotes the gating variable for subunit activation, the total potassium conductance is proportional to $n^{4}$.

This four-subunits interpretation suggests molecular-scale conformational changes that exceed the accuracy of the Hodgkin-Huxley model. However, this concept provides a theoretical interpretation of the fundamental nonlinearity of the neuron that is interesting for exploring equivalent forms of the Hodgkin-Huxley model.

The original Hodgkin-Huxley model is nonlinear in two ways: one is the use of an exponentiation for the gating variable, such as $n^{4}$ ; the other is the voltage dependence of the rate constants. In the first case, an exponentiation is used to describe the delay in current response observed after a step change in membrane potential. In the second case, the use of the voltage clamp at a constant membrane potential $V_{0}$ has the effect of linearizing the differential equation $\frac{dn}{dt}=\alpha_{n}\left(V_{0}\right)\left(1-n\right)-\beta_{n}\left(V_{0}\right)n$ since the rate constants $\alpha_{n}\left(V_{0}\right)$ and $\beta_{n}\left(V_{0}\right)$ become constants.

The two-state process can be extended to multiple sequential states typical of higher order chemical relaxation models. More precisely, the four independent identical subunits of the potassium ion channel can be modeled as a five-state Markov chain
\begin{equation}
0\overset{4\alpha_{n}}{\underset{\beta_{n}}{\rightleftharpoons}}1\overset{3\alpha_{n}}{\underset{2\beta_{n}}{\rightleftharpoons}}2\overset{2\alpha_{n}}{\underset{3\beta_{n}}{\rightleftharpoons}}3\overset{\alpha_{n}}{\underset{4\beta_{n}}{\rightleftharpoons}}4
\label{eq:n4_markov_chain}
\end{equation}
where each state $0$, $1$, $2$, $3$, $4$ corresponds to the number of open subunits at a given time. Each channel has the probability $p_{k}\left(t\right)$ to be in state $k$. In particular, $1=p_{0}+p_{1}+p_{2}+p_{3}+p_{4}$. The channel is open when all four subunits are open, which corresponds to state $4$ with probability $p_{4}$. When the system is in the state $k$, there are $k$ open subunits and $\left(4-k\right)$ closed subunits. For $k<4$, the transition $k\rightarrow k+1$ opens one of the $\left(4-k\right)$ closed subunits, which corresponds to the rate $\left(4-k\right)\alpha_{n}$. Similarly, for $k>0$, the transition $k\rightarrow k-1$ closes one of the $k$ open subunits, which corresponds to the rate $k\beta_{n}$.

More generally, arbitrary rate constants could be chosen other than integer multiples of $\alpha_{n}$ and $\beta_{n}$. Numerous models of this type with more than two independent rate constants have been compared to the original Hodgkin-Huxley model, e.g. \cite{Vandenberg1991}.

Stochastic Markov simulations were based on \cite{Gillespie1977} and \cite{Goldwyn2011a,Goldwyn2011b} considering a statistical population of $N$ potassium ion channels and a time interval $t\in\left[0;T\right]$ discretized by $\Delta t$. Let $N_{k}\left(t\right)$ be the number of channels in state $k$ at time $t$. In particular, $N=N_{0}+N_{1}+N_{2}+N_{3}+N_{4}$. In the limit, for a large population, the proportion of channels in state $k$ tends to the corresponding probability
\[
\lim_{N\rightarrow\infty}\frac{N_{k}\left(t\right)}{N}=p_{k}\left(t\right)
\]
Let $\Delta N_{k}\left(t\right)$ be the variation of the number of channels in state $k$ during $\Delta t$. Let $\Delta N_{i\rightarrow j}\left(t\right)$ be the number of channels switching from state $i$ to state $j$ during $\Delta t$.

To ensure validity, it will always be assumed that $p_{k}$, $N_{k}$, $\Delta N_{k}$, $\Delta N_{i\rightarrow j}$ are zero for indices outside the range $\left\{0,1,2,3,4\right\}$.

The proportion of channels switching from state $k$ to state $k\pm1$ during $\Delta t$ is determined by the transition rates
\begin{eqnarray*}
	\Delta N_{k\rightarrow k+1} & = & \left(4-k\right)\alpha_{n}N_{k}\Delta t\\*
	\Delta N_{k\rightarrow k-1} & = & k\beta_{n}N_{k}\Delta t
\end{eqnarray*}

Furthermore, the variation of the number of channels in state $k$ during $\Delta t$ corresponds to the number of transitions to state $k$ minus the number of transitions from state $k$. More precisely
\[
\Delta N_{k}=\Delta N_{k-1\rightarrow k}+\Delta N_{k+1\rightarrow k}-\Delta N_{k\rightarrow k-1}-\Delta N_{k\rightarrow k+1}
\]

Replacing expressions
\[
\Delta N_{k}=\left(5-k\right)\alpha_{n}N_{k-1}\Delta t+\left(k+1\right)\beta_{n}N_{k+1}\Delta t-k\beta_{n}N_{k}\Delta t-\left(4-k\right)\alpha_{n}N_{k}\Delta t
\]

Dividing by $N$ and taking the limit, a system of differential equations is obtained
\[
\frac{dp_{k}}{dt}=\left(5-k\right)\alpha_{n}p_{k-1}+\left(k+1\right)\beta_{n}p_{k+1}-k\beta_{n}p_{k}-\left(4-k\right)\alpha_{n}p_{k}
\]

More explicitly, this gives the master equation
\begin{eqnarray*}
	\frac{dp_{0}}{dt} & = & -4\alpha_{n}p_{0}+\beta_{n}p_{1}\\*
	\frac{dp_{1}}{dt} & = & 4\alpha_{n}p_{0}-\left(3\alpha_{n}+\beta_{n}\right)p_{1}+2\beta_{n}p_{2}\\*
	\frac{dp_{2}}{dt} & = & 3\alpha_{n}p_{1}-2\left(\alpha_{n}+\beta_{n}\right)p_{2}+3\beta_{n}p_{3}\\*
	\frac{dp_{3}}{dt} & = & 2\alpha_{n}p_{2}-\left(\alpha_{n}+3\beta_{n}\right)p_{3}+4\beta_{n}p_{4}\\*
	\frac{dp_{4}}{dt} & = & \alpha_{n}p_{3}-4\beta_{n}p_{4}
\end{eqnarray*}

The conductance of the population of channels is calculated as $g_{\mathrm{K}}f$ where $g_{\mathrm{K}}$ is the conductance of an individual channel and $f$ is the proportion of open channels \citep{Goldwyn2011a}. In the limit, for a large population, the proportion of open channels tends to the probability $p_{4}$ that a channel is open, which corresponds to the deterministic conductance $g_{\mathrm{K}}p_{4}$.

Although the master equation does not use exponentiation, the equivalence with the $n^{4}$ model is discussed by \cite{Dayan2001}. Indeed, in the Hodgkin-Huxley model, $n$ is a gating variable representing the probability that a subunit is open and $1-n$ the probability that it is closed. In state $k$, $k$ of the four subunits are open and the $4-k$ others are closed, thus
$p_{k}=\left(\begin{array}{c} 4\\ k \end{array}\right)n^{k}\left(1-n\right)^{4-k}$. In particular, in state $4$, all four subunits are open, thus $p_{4}=n^{4}$ which is consistent with the exponentiation of the Hodgkin-Huxley model.

Therefore, at a fundamental level, nonlinearities generated by neuronal responses are likely to reflect fluctuations between internal states for which the master equation is controlled by transition rates involving energy for the movement of charges, the associated Boltzmann factor is discussed by \cite{Dayan2001}. The master equation provides a deterministic description of the probabilities of these fluctuations.

\subsection{Markov power spectra}\label{subsec_markov_power_spectra}

Stochastic analysis of Markov models provides power spectra of conductance noise identical to the first interpretation of \cite{Stevens_1972} involving probability and correlation functions. Many papers have derived noise power spectra based on the nonlinear nature of the potassium channel ($n^{4}$), especially \cite{ODonnell2014} which is used in this article. In addition to the squid axon, measurements on the nodes of Ranvier \citep{Conti1984} are consistent with the nonlinear origin of the conductance noise power spectra. Such an interpretation that certain nonlinear properties can induce a probabilistic or stochastic character, suggests that they are involved in the fundamental origin of spontaneous fluctuations in neurons.

The conductance noise can be predicted from the Hodgkin-Huxley equations for potassium conductance as described by \cite{ODonnell2014}. Their approach is explored here for further adaptation in this article.

Let $X_{t}$ be the random variable measuring if the channel is open ($X_{t}=1$) or closed ($X_{t}=0$) at time $t$. Let $p\left(t\right)$ be the probability that the channel is open at time $t$, namely $p\left(t\right)=p\left(X_{t}=1\right)$. Let $p_{\infty}$ be the steady state probability, which coincides with the average of $p\left(t\right)$ over time. Actually, $p=p_{4}=n^{4}$ using the previous Markov probability notations, but the $p$ notation is used here to be more general.

By definition, the autocorrelation $r\left(t_{1},t_{2}\right)$ characterizes the similarity of $X_{t}$ between times $t_{1}$ and $t_{2}$. Denoting by $\mathbf{E}$ the expected value of a random variable
\[
r\left(t_{1},t_{2}\right)=\mathbf{E}\left[X_{t_{1}}X_{t_{2}}\right]
\]
Since $X_{t}$ only takes the values $0$ or $1$, the autocorrelation is given by
\begin{eqnarray*}
	r\left(t_{1},t_{2}\right) & = & \sum_{i,j\in\left\{ 0,1\right\} }i\cdot j\cdot p\left(X_{t_{1}}=i,X_{t_{2}}=j\right)\\*
	& = & p\left(X_{t_{1}}=1,X_{t_{2}}=1\right)\\*
	& = & p\left(X_{t_{2}}=1\vert X_{t_{1}}=1\right)p\left(X_{t_{1}}=1\right)
\end{eqnarray*}
or more concisely
\[
r\left(t_{1},t_{2}\right)=p\left(t_{1}\right)p\left(t_{2}\vert t_{1}\right)
\]

The autocorrelation can also be reformulated betwen times $t_{0}$ and $t_{0}+s$ to make the time lag $s$ explicit
\[
r\left(t_{0},t_{0}+s\right)=p\left(t_{0}\right)p\left(t_{0}+s\vert t_{0}\right)
\]
Here, $p\left(t_{0}+s\vert t_{0}\right)$ is the conditional probability that the channel is open at time $t_{0}+s$ provided that it is open at time $t_{0}$. Importantly, if the channel is open at time $t_{0}$ then $p\left(t_{0}\right)=1$. In this special case, the Markov state $\left(p_{0},p_{1},p_{2},p_{3},p_{4}\right)=\left(0,0,0,0,1\right)$ is unique due to the constraint $p_{0}+p_{1}+p_{2}+p_{3}+p_{4}=1$. The same reasoning is valid for other sequential models with fewer or more Markov states. Thus, the time evolution of $p\left(t_{0}+s\vert t_{0}\right)$ from the unique state $\left(0,0,0,0,1\right)$ between $t_{0}$ and $t_{0}+s$ is identical to the time evolution of $p\left(s\vert0\right)$ from the unique state $\left(0,0,0,0,1\right)$ between $0$ and $s$, the two trajectories of the dynamical system are identical because they start from the same unique state $\left(0,0,0,0,1\right)$ and have the same duration $s$. Therefore
\[
r\left(t_{0},t_{0}+s\right)=p\left(t_{0}\right)p\left(s\vert0\right)
\]

Averaging over the time origin $t_{0}$ gives the autocorrelation as a function of the time lag
\begin{eqnarray*}
	r\left(s\right) & = & \langle r\left(t_{0},t_{0}+s\right)\rangle_{t_{0}}\\*
	& = & \langle p\left(t_{0}\right)p\left(s\vert0\right)\rangle_{t_{0}}\\*
	& = & \langle p\left(t_{0}\right)\rangle_{t_{0}}p\left(s\vert0\right)\\*
	& = & p_{\infty}p\left(s\vert0\right)
\end{eqnarray*}

By definition, the autocovariance $C\left(t_{1},t_{2}\right)$ characterizes the similarity of $X_{t}-\mathbf{E}\left[X_{t}\right]$ between times $t_{1}$ and $t_{2}$. Namely
\begin{eqnarray*}
	C\left(t_{1},t_{2}\right) & = & \mathbf{E}\left[\left(X_{t_{1}}-\mathbf{E}\left[X_{t_{1}}\right]\right)\left(X_{t_{2}}-\mathbf{E}\left[X_{t_{2}}\right]\right)\right]\\*
	& = & \mathbf{E}\left[\left(X_{t_{1}}-p_{\infty}\right)\left(X_{t_{2}}-p_{\infty}\right)\right]\\*
	& = & \mathbf{E}\left[X_{t_{1}}X_{t_{2}}\right]+\mathbf{E}\left[-p_{\infty}X_{t_{1}}\right]+\mathbf{E}\left[-p_{\infty}X_{t_{2}}\right]+\mathbf{E}\left[p_{\infty}^{2}\right]\\*
	& = & r\left(t_{1},t_{2}\right)-p_{\infty}\mathbf{E}\left[X_{t_{1}}\right]-p_{\infty}\mathbf{E}\left[X_{t_{2}}\right]+p_{\infty}^{2}\\*
	& = & r\left(t_{1},t_{2}\right)-p_{\infty}^{2}-p_{\infty}^{2}+p_{\infty}^{2}\\*
	& = & r\left(t_{1},t_{2}\right)-p_{\infty}^{2}
\end{eqnarray*}

The autocovariance can also be reformulated betwen times $t_{0}$ and $t_{0}+s$ to make the time lag $s$ explicit
\[
C\left(t_{0},t_{0}+s\right)=r\left(t_{0},t_{0}+s\right)-p_{\infty}^{2}
\]

Averaging over the time origin $t_{0}$ gives the autocovariance as a function of the time lag
\begin{eqnarray*}
	C\left(s\right) & = & \langle C\left(t_{0},t_{0}+s\right)\rangle_{t_{0}}\\*
	& = & r\left(s\right)-p_{\infty}^{2}\\*
	& = & p_{\infty}p\left(s\vert0\right)-p_{\infty}^{2}
\end{eqnarray*}

When the channel is open, the single-channel current is given by Ohm's law, namely $i_{\mathrm{K}}=\gamma_{\mathrm{K}}\left(V-V_{\mathrm{K}}\right)$ where $\gamma_{K}$ is the single-channel conductance, $V$ is the voltage imposed by voltage clamp and $V_{\mathrm{K}}$ is the reversal potential. Then, the fluctuating single-channel current is defined as $i_{\mathrm{K}}\cdot X_{t}$. In particular, the fluctuating single-channel current is zero when the channel is closed and is given by Ohm's law when the channel is open. For a population of $N_{\mathrm{K}}$ channels, let $X_{t}^{n}$ be the random variables for each channel $n=1\ldots N_{\mathrm{K}}$. The total fluctuating current is
\[
I_{\mathrm{K}}\left(t\right)=\sum_{n=1}^{N_{\mathrm{K}}}i_{\mathrm{K}}\cdot X_{t}^{n}
\]

By definition, the autocovariance $C_{I_{\mathrm{K}}}\left(t_{1},t_{2}\right)$ of the total fluctuating current between times $t_{1}$ and $t_{2}$ is given by
\begin{eqnarray*}
	C_{I_{\mathrm{K}}}\left(t_{1},t_{2}\right) & = & \mathbf{E}\left[\left(I_{\mathrm{K}}\left(t_{1}\right)-\mathbf{E}\left[I_{\mathrm{K}}\left(t_{1}\right)\right]\right)\left(I_{\mathrm{K}}\left(t_{2}\right)-\mathbf{E}\left[I_{\mathrm{K}}\left(t_{2}\right)\right]\right)\right]\\*
	& = & \mathbf{E}\left[\left(\sum_{n=1}^{N_{\mathrm{K}}}i_{\mathrm{K}}\cdot X_{t_{1}}^{n}-\mathbf{E}\left[\sum_{n=1}^{N_{\mathrm{K}}}i_{\mathrm{K}}\cdot X_{t_{1}}^{n}\right]\right)\left(\sum_{m=1}^{N_{\mathrm{K}}}i_{\mathrm{K}}\cdot X_{t_{2}}^{m}-\mathbf{E}\left[\sum_{m=1}^{N_{\mathrm{K}}}i_{\mathrm{K}}\cdot X_{t_{2}}^{m}\right]\right)\right]\\*
	& = & i_{\mathrm{K}}^{2}\mathbf{E}\left[\left(\sum_{n=1}^{N_{\mathrm{K}}}X_{t_{1}}^{n}-\sum_{n=1}^{N_{\mathrm{K}}}\mathbf{E}\left[X_{t_{1}}^{n}\right]\right)\left(\sum_{m=1}^{N_{\mathrm{K}}}X_{t_{2}}^{m}-\sum_{m=1}^{N_{\mathrm{K}}}\mathbf{E}\left[X_{t_{2}}^{m}\right]\right)\right]\\*
	& = & i_{\mathrm{K}}^{2}\mathbf{E}\left[\left(\sum_{n=1}^{N_{\mathrm{K}}}\left(X_{t_{1}}^{n}-\mathbf{E}\left[X_{t_{1}}^{n}\right]\right)\right)\left(\sum_{m=1}^{N_{\mathrm{K}}}\left(X_{t_{2}}^{m}-\mathbf{E}\left[X_{t_{2}}^{m}\right]\right)\right)\right]\\*
	& = & i_{\mathrm{K}}^{2}\sum_{n=1}^{N_{\mathrm{K}}}\sum_{m=1}^{N_{\mathrm{K}}}\mathbf{E}\left[\left(X_{t_{1}}^{n}-p_{\infty}\right)\left(X_{t_{2}}^{m}-p_{\infty}\right)\right]
\end{eqnarray*}

It can be noticed that
\begin{eqnarray*}
	\mathbf{E}\left[\left(X_{t_{1}}^{n}-p_{\infty}\right)\left(X_{t_{2}}^{m}-p_{\infty}\right)\right] & = & \mathbf{E}\left[X_{t_{1}}^{n}X_{t_{2}}^{m}-p_{\infty}X_{t_{1}}^{n}-p_{\infty}X_{t_{2}}^{m}+p_{\infty}^{2}\right]\\*
	& = & \mathbf{E}\left[X_{t_{1}}^{n}X_{t_{2}}^{m}\right]-p_{\infty}\mathbf{E}\left[X_{t_{1}}^{n}\right]-p_{\infty}\mathbf{E}\left[X_{t_{2}}^{m}\right]+\mathbf{E}\left[p_{\infty}^{2}\right]\\*
	& = & \mathbf{E}\left[X_{t_{1}}^{n}X_{t_{2}}^{m}\right]-p_{\infty}^{2}-p_{\infty}^{2}+p_{\infty}^{2}\\*
	& = & \mathbf{E}\left[X_{t_{1}}^{n}X_{t_{2}}^{m}\right]-p_{\infty}^{2}
\end{eqnarray*}

If $n\neq m$, the random variables $X_{t}^{n}$ and $X_{t}^{m}$ are independent, then
\[
\mathbf{E}\left[\left(X_{t_{1}}^{n}-p_{\infty}\right)\left(X_{t_{2}}^{m}-p_{\infty}\right)\right]=\mathbf{E}\left[X_{t_{1}}^{n}\right]\mathbf{E}\left[X_{t_{2}}^{m}\right]-p_{\infty}^{2}=p_{\infty}p_{\infty}-p_{\infty}^{2}=0
\]

If $n=m$, the random variables $X_{t}^{n}$ and $X_{t}^{m}$ are the same as $X_{t}$, then
\[
\mathbf{E}\left[\left(X_{t_{1}}^{n}-p_{\infty}\right)\left(X_{t_{2}}^{m}-p_{\infty}\right)\right]=\mathbf{E}\left[X_{t_{1}}X_{t_{2}}\right]-p_{\infty}^{2}=r\left(t_{1},t_{2}\right)-p_{\infty}^{2}=C\left(t_{1},t_{2}\right)
\]

These remarks on the indices $n$ and $m$ allow to simplify the double sum in the autocovariance of the total fluctuating current so that
\begin{eqnarray*}
	C_{I_{\mathrm{K}}}\left(t_{1},t_{2}\right) & = & i_{\mathrm{K}}^{2}\sum_{n=1}^{N_{\mathrm{K}}}C\left(t_{1},t_{2}\right)\\*
	& = & N_{\mathrm{K}}i_{\mathrm{K}}^{2}C\left(t_{1},t_{2}\right)
\end{eqnarray*}

The autocovariance can also be reformulated betwen times $t_{0}$
and $t_{0}+s$ to make the time lag $s$ explicit
\[
C_{I_{\mathrm{K}}}\left(t_{0},t_{0}+s\right)=N_{\mathrm{K}}i_{\mathrm{K}}^{2}C\left(t_{0},t_{0}+s\right)
\]

Averaging over the time origin $t_{0}$ gives the autocovariance as
a function of the time lag
\begin{eqnarray*}
	C_{I_{\mathrm{K}}}\left(s\right) & = & \langle N_{\mathrm{K}}i_{\mathrm{K}}^{2}C\left(t_{0},t_{0}+s\right)\rangle_{t_{0}}\\*
	& = & N_{\mathrm{K}}i_{\mathrm{K}}^{2}\langle C\left(t_{0},t_{0}+s\right)\rangle_{t_{0}}\\*
	& = & N_{\mathrm{K}}i_{\mathrm{K}}^{2}C\left(s\right)
\end{eqnarray*}

The Wiener-Khinchin theorem provides the power spectrum $S_{I_{\mathrm{K}}}^{\pm}$ as the Fourier transform of the autocovariance $C_{I_{\mathrm{K}}}$, using the non-unitary Fourier transform with angular frequencies
\[
S_{I_{\mathrm{K}}}^{\pm}\left(\omega\right)=\int_{-\infty}^{+\infty}C_{I_{\mathrm{K}}}\left(s\right)e^{-i\omega s}ds
\]

The autocovariance is an even function of the lag because under stationarity, the average over the time origin $t_{0}$ is independent of a time lag
\begin{eqnarray*}
	C\left(s\right) & = & \langle C\left(t_{0},t_{0}+s\right)\rangle_{t_{0}}\\*
	& = & \langle C\left(t_{0}-s,t_{0}\right)\rangle_{t_{0}-s}\\*
	& = & \langle C\left(t_{0}-s,t_{0}\right)\rangle_{t_{0}}\\*
	& = & \langle C\left(t_{0},t_{0}-s\right)\rangle_{t_{0}}\\*
	& = & C\left(-s\right)
\end{eqnarray*}

Thus, the power spectrum can be decomposed into two parts
\begin{eqnarray*}
	S_{I_{\mathrm{K}}}^{\pm}\left(\omega\right) & = & \int_{-\infty}^{0}C_{I_{\mathrm{K}}}\left(s\right)e^{-i\omega s}ds+\int_{0}^{+\infty}C_{I_{\mathrm{K}}}\left(s\right)e^{-i\omega s}ds\\*
	& = & \int_{0}^{+\infty}C_{I_{\mathrm{K}}}\left(-s\right)e^{i\omega s}ds+\int_{0}^{+\infty}C_{I_{\mathrm{K}}}\left(s\right)e^{-i\omega s}ds\\*
	& = & \int_{0}^{+\infty}C_{I_{\mathrm{K}}}\left(s\right)\left[e^{i\omega s}+e^{-i\omega s}\right]ds\\*
	& = & 2\int_{0}^{+\infty}C_{I_{\mathrm{K}}}\left(s\right)\cos\left(\omega s\right)ds
\end{eqnarray*}

To follow the convention of \cite{ODonnell2014}, the power spectrum $S_{I_{\mathrm{K}}}$ will be considered for positive frequencies only, i.e.
\[
S_{I_{\mathrm{K}}}\left(\omega\right)=2S_{I_{\mathrm{K}}}^{\pm}\left(\omega\right)=4\Re\int_{0}^{+\infty}C_{I_{\mathrm{K}}}\left(s\right)e^{-i\omega s}ds
\]

In order to calculate the Markov power spectra for the fluctuating potassium current, it is necessary to determine the autocovariance $C\left(s\right)=p_{\infty}p\left(s\vert0\right)-p_{\infty}^{2}$ based on the two components $p\left(s\vert0\right)$ and $p_{\infty}$. The kinetics of the potassium gating variable $n$ is given by
\[
\frac{dn}{dt}=\alpha_{n}\left(1-n\right)-\beta_{n}n
\]
or equivalently
\[
\frac{dn}{dt}=\frac{n_{\infty}-n}{\tau_{n}}
\]

The general solution can be formulated with an exponential decay and an arbitrary initial condition $n_{0}$
\[
n\left(t\right)=n_{\infty}+\left(n_{0}-n_{\infty}\right)e^{-t/\tau_{n}}
\]

The solution $n\left(t\vert0\right)$ is interpreted as the conditional probability that the gate is open at time $t$ provided that it is open at time $0$, which is implemented by the initial condition $n_{0}=1$
\[
n\left(t|0\right)=n_{\infty}+\left(1-n_{\infty}\right)e^{-t/\tau_{n}}
\]

The potassium ion channel having four independent identical subunits, this provides the probability that the channel is open at time $t$
\[
p\left(t\right)=n^{4}\left(t\right)
\]

At steady state
\[
p_{\infty}=n_{\infty}^{4}
\]

The conditional probability that the channel is open at time $t$ provided that the channel is open at time $t=0$ is given by
\begin{eqnarray*}
	p\left(t\vert0\right) & = & n^{4}\left(t\vert0\right)\\*
	& = & \left[n_{\infty}+\left(1-n_{\infty}\right)e^{-t/\tau_{n}}\right]^{4}\\*
	& = & \sum_{q=0}^{4}\binom{4}{q}n_{\infty}^{4-q}\left(1-n_{\infty}\right)^{q}e^{-q t/\tau_{n}}
\end{eqnarray*}

The autocovariance does follow
\begin{eqnarray*}
	C_{I_{\mathrm{K}}}\left(t\right) & = & N_{\mathrm{K}}i_{\mathrm{K}}^{2}C\left(t\right)\\*
	& = & N_{\mathrm{K}}i_{\mathrm{K}}^{2}\left[p_{\infty}p\left(t\vert0\right)-p_{\infty}^{2}\right]\\*
	& = & N_{\mathrm{K}}i_{\mathrm{K}}^{2}\left[n_{\infty}^{4}p\left(t\vert0\right)-\left(n_{\infty}^{4}\right)^{2}\right]\\*
	& = & N_{\mathrm{K}}i_{\mathrm{K}}^{2}n_{\infty}^{4}\left[p\left(t\vert0\right)-n_{\infty}^{4}\right]\\*
	& = & N_{\mathrm{K}}i_{\mathrm{K}}^{2}n_{\infty}^{4}\sum_{q=1}^{4}\binom{4}{q}n_{\infty}^{4-q}\left(1-n_{\infty}\right)^{q}e^{-q t/\tau_{n}}
\end{eqnarray*}
where the term corresponding to $q=0$ in the sum has been canceled with $-n_{\infty}^{4}$.

The power spectrum $S_{I_{\mathrm{K}}}$ is computed by integrating each term of the sum in $C_{I_{\mathrm{K}}}$
\begin{eqnarray*}
	S_{q}\left(\omega\right) & = & \int_{0}^{+\infty}\binom{4}{q}n_{\infty}^{4-q}\left(1-n_{\infty}\right)^{q}e^{-qt/\tau_{n}}e^{-i\omega t}dt\\*
	& = & \binom{4}{q}n_{\infty}^{4-q}\left(1-n_{\infty}\right)^{q}\int_{0}^{+\infty}e^{-qt/\tau_{n}}e^{-i\omega t}dt
\end{eqnarray*}

Using the non-unitary Fourier transform $\mathcal{F}\left[e^{-a\left|t\right|}\right]=\frac{2a}{a^{2}+\omega^{2}}$ with angular frequencies and divided by $2$ because of the positive half of the time $\left[0;+\infty\right]$
\begin{eqnarray*}
	S_{q}\left(\omega\right)& = & \binom{4}{q}n_{\infty}^{4-q}\left(1-n_{\infty}\right)^{q}\frac{q/\tau_{n}}{q^{2}/\tau_{n}^{2}+\omega^{2}}\\*
	& = & \binom{4}{q}n_{\infty}^{4-q}\left(1-n_{\infty}\right)^{q}\frac{q\tau_{n}}{q^{2}+\omega^{2}\tau_{n}^{2}}
\end{eqnarray*}

Applying $S_{q}$ to each $q$
\begin{eqnarray*}
	S_{1}\left(\omega\right) & = & 4n_{\infty}^{3}\left(1-n_{\infty}\right)\frac{\tau_{n}}{1+\omega^{2}\tau_{n}^{2}}\\*
	S_{2}\left(\omega\right) & = & 6n_{\infty}^{2}\left(1-n_{\infty}\right)^{2}\frac{2\tau_{n}}{4+\omega^{2}\tau_{n}^{2}}\\*
	S_{3}\left(\omega\right) & = & 4n_{\infty}\left(1-n_{\infty}\right)^{3}\frac{3\tau_{n}}{9+\omega^{2}\tau_{n}^{2}}\\*
	S_{4}\left(\omega\right) & = & \left(1-n_{\infty}\right)^{4}\frac{4\tau_{n}}{16+\omega^{2}\tau_{n}^{2}}
\end{eqnarray*}
the power spectrum is obtained
\begin{equation}
	S_{I_{\mathrm{K}}}\left(\omega\right)=4N_{\mathrm{K}}i_{\mathrm{K}}^{2}n_{\infty}^{4}\left[S_{1}\left(\omega\right)+S_{2}\left(\omega\right)+S_{3}\left(\omega\right)+S_{4}\left(\omega\right)\right]\label{eq:rossum_power_spectrum_n4}
\end{equation}

There are four Lorentzians with corner frequencies $\omega_{1}=\frac{1}{\tau_{n}}$, $\omega_{2}=\frac{2}{\tau_{n}}$, $\omega_{3}=\frac{3}{\tau_{n}}$, $\omega_{4}=\frac{4}{\tau_{n}}$ respectively. In particular, $j\omega_{q}$ correspond to the poles.

The parameters of Table \ref{tab_linear_k_na_parameters} were reused with the potassium channel density $\rho_{\mathrm{K}}=18/\mathrm{\mu m}^{2}$, the membrane area $A_{\mathrm{K}}=500\mathrm{\mu m}^{2}$, the number of potassium channels $N_{\mathrm{K}}=\rho_{\mathrm{K}}A_{\mathrm{K}}$ and the potassium single-channel conductance $\gamma_{\mathrm{K}}=g_{\mathrm{K}}/N_{\mathrm{K}}$.

\begin{figure}[H]
\centering{}
\includegraphics[width=12cm]{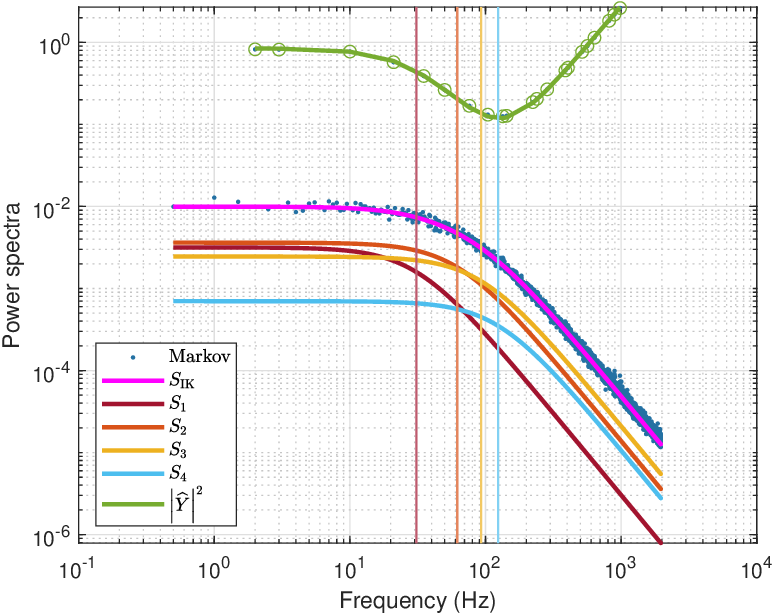}
\caption{\label{fig:qsa_vc_n4_5mV_M128_noise_stimulated}\textbf{Markov simulation of the $n^{4}$ model for a $5\mathrm{mV}$ depolarization with stimulus amplitude $0.25\mathrm{mV}$.} The power spectra approximated by $128$ iterations are represented by the blue scatterplot. The predicted power spectrum $S_{\mathrm{IK}}$ (magenta curve) accurately fits the blue scatterplot. The four Lorentzians compose the predicted power spectrum with $S_{1}$ (brown curve), $S_{2}$ (orange curve), $S_{3}$ (yellow curve), $S_{4}$ (blue curve). The corresponding corner frequencies $\omega_{1}$, $\omega_{2}$, $\omega_{3}$, $\omega_{4}$ are represented by vertical lines. The squared admittance $\left|\widehat{Y}\right|^{2}$ (green curve) matches the linear Markov responses after adjustment of the vertical offset. The QSA multi-sinusoidal stimulus frequencies are {[}2, 3, 10, 21, 35, 50, 76, 104, 134, 143, 223, 239, 285, 388,405, 515, 564, 636, 815, 892, 982{]} Hertz.}
\end{figure}

\begin{figure}[H]
\centering{}
\includegraphics[width=12cm]{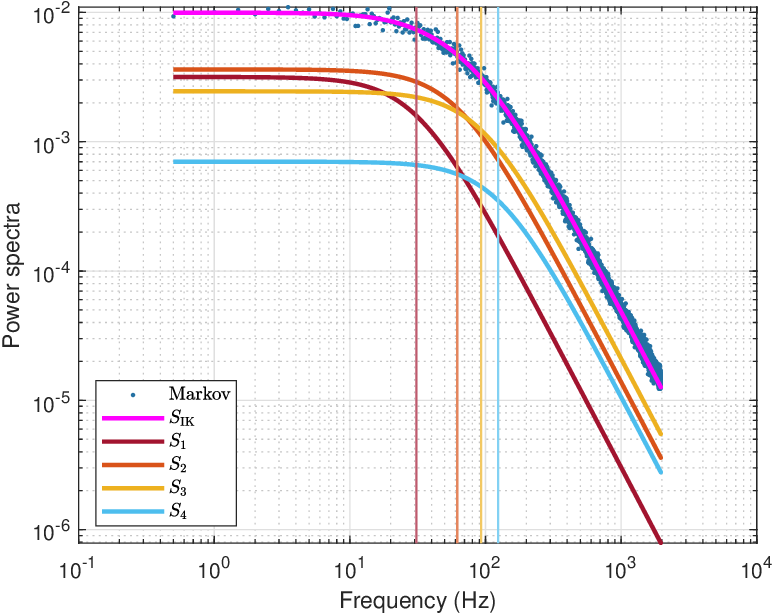}
\caption{\label{fig:qsa_vc_n4_5mV_M128_noise_notstimulated}\textbf{Markov simulation of the $n^{4}$ model for a $5\mathrm{mV}$ depolarization without stimulus.} The fluctuation power spectra are the same as those in Fig. \ref{fig:qsa_vc_n4_5mV_M128_noise_stimulated} but there is no admittance.}
\end{figure}

Fig. \ref{fig:qsa_vc_n4_5mV_M128_noise_stimulated} illustrates a voltage clamp Markov simulation (Eq. \ref{eq:n4_markov_chain}) for a $5\mathrm{mV}$ depolarization of the spontaneous potassium current fluctuations superimposed on the theoretically predicted power spectrum $S_{\mathrm{IK}}$, the four Lorentzians, the four corner frequencies and the admittance. The simulation was done with the Markov model and QSA multi-sinusoidal stimuli of amplitude $0.25\mathrm{mV}$. At this potential, the predicted power spectrum $S_{\mathrm{IK}}$ (magenta curve) have relative equivalent contributions from all the Lorentzian functions $S_{1}$, $S_{2}$, $S_{3}$, $S_{4}$ and certainly not just those present in the lowest corner frequency, $\omega_{4}=\frac{1}{\tau_{n}}$. The voltage responses at the stimulus frequencies are accurately described by the squared admittance $\left|\widehat{Y}\right|^{2}$ (green curve) after adjustment of the vertical offset. The square of the admittance is obtained with an input of constant amplitude, which makes it possible to compare it to other curves independent of the input. Fig. \ref{fig:qsa_vc_n4_5mV_M128_noise_notstimulated} shows that simulation done without stimulus gave identical power spectra.

\section{Results}\label{sec_results}

\subsection{The $n^{2}$ model}\label{subsec_n2_model}

Since chemical relaxation models generally obey the fluctuation-dissipation theorem (FDT), it is useful to compare their linear and nonlinear behaviors using both averaged quadratic (QSA) and noise (Markov) power spectra. This will be done with the potassium conductance of Hodgkin-Huxley equations, namely $n^{4}$, and compared with more general relaxation models having fewer sequential states. A similar comparison could be done with the sodium conductance.

A variant of the Hodgkin-Huxley equations, the \cite{Frankenhaeuser1964} equations for myelinated nerve, uses a $n^{2}$ model for the potassium conductance, which can be simulated as a three-state Markov chain
\[
0\overset{2\alpha_{n}}{\underset{\beta_{n}}{\rightleftharpoons}}1\overset{\alpha_{n}}{\underset{2\beta_{n}}{\rightleftharpoons}}2
\]

Thus, the $n^{2}$ model has only two independent rate constants, $\alpha_{n}$ and $\beta_{n}$. More generally, such a $n^{2}$ model is a particular sequential kinetic model $0\rightleftharpoons1\rightleftharpoons2$, arbitrarily called $p_{2}$ in this paper, having the four specific rate constants above with only two independent rate constants. However, a general $p_{2}$ model, which has four independent rate constants, will be shown to be a reasonable model for a potassium conductance. In particular, the rate constants of the $p_{2}$ model can be selected to show similar behavior to the Hodgkin-Huxley $n^{4}$ model. In this case, the $p_{2}$ model is a model without exponentiation, which is not identical to the $n^{2}$ model, but approximates the Hodgkin-Huxley $n^{4}$ model.

\subsection{The $p_{2}$ model}\label{subsec_p2_model}

The general $p_{2}$ model has four independent rate constants, two forward $k_{1}$, $k_{3}$, and two backward $k_{4}$, $k_{2}$, as follows:
\begin{equation}
p_{0}\overset{k_{1}}{\underset{k_{2}}{\rightleftharpoons}}p_{1}\overset{k_{3}}{\underset{k_{4}}{\rightleftharpoons}}p_{2}
\label{eq:p2_markov_chain}
\end{equation}

The $n^{2}$ model is a special case of the $p_{2}$ model with $k_{1}=2\alpha_{n}$, $k_{3}=\alpha_{n}$, $k_{4}=2\beta_{n}$ and $k_{2}=\beta_{n}$.

More generally, it may be convenient to rewrite $k_{1}$, $k_{2}$, $k_{3}$, $k_{4}$ in terms of $\alpha_{n}$ and $\beta_{n}$ such that $k_{1}=A\alpha_{n}$, $k_{3}=\alpha_{n}$, $k_{4}=B\beta_{n}$ and $k_{2}=\beta_{n}$, where $A$ and $B$ are arbitrary factors. In this way, the $n^{2}$ model is a special case of the $p_{2}$ model with $A=B=2$.

In this article, the $p_{2}$ model simulations use $A=0.35$ and $B=4$. The other parameters are the same as the $n^{4}$ model simulations given in Table \ref{tab_linear_k_na_parameters}.

The master equation is deduced from the three-state Markov chain as follows
\begin{eqnarray*}
	\frac{dp_{0}}{dt} & = & -k_{1}p_{0}+k_{2}p_{1}\\*
	\frac{dp_{1}}{dt} & = & k_{1}p_{0}-\left(k_{2}+k_{3}\right)p_{1}+k_{4}p_{2}\\*
	\frac{dp_{2}}{dt} & = & k_{3}p_{1}-k_{4}p_{2}
\end{eqnarray*}

Since $p_{0}+p_{1}+p_{2}=1$, the system can be reduced to the variables $p_{1}$ and $p_{2}$
\begin{eqnarray}
	\frac{dp_{1}}{dt} & = & -\left(k_{1}+k_{2}+k_{3}\right)p_{1}+\left(-k_{1}+k_{4}\right)p_{2}+k_{1}\label{eq:p2_ode}\\*
	\frac{dp_{2}}{dt} & = & k_{3}p_{1}-k_{4}p_{2}\nonumber 
\end{eqnarray}

This can be rewritten in matrix form
\begin{equation}
	\left(\begin{array}{c}
		\frac{dp_{1}}{dt}\\
		\frac{dp_{2}}{dt}
	\end{array}\right)=\left(\begin{array}{cc}
		\alpha_{11} & \alpha_{12}\\
		\alpha_{21} & \alpha_{22}
	\end{array}\right)\left(\begin{array}{c}
		p_{1}\\
		p_{2}
	\end{array}\right)+\left(\begin{array}{c}
		-\alpha_{22}-\alpha_{12}\\
		0
	\end{array}\right)\label{eq:p2_ode_matrix}
\end{equation}
where
\begin{eqnarray*}
	\alpha_{11} & = & -\left(k_{1}+k_{2}+k_{3}\right)\\*
	\alpha_{12} & = & -k_{1}+k_{4}\\*
	\alpha_{21} & = & k_{3}\\*
	\alpha_{22} & = & -k_{4}
\end{eqnarray*}

At steady state
\begin{eqnarray}
	\left(\begin{array}{c}
		p_{1}^{\infty}\\
		p_{2}^{\infty}
	\end{array}\right) & = & \left(\begin{array}{cc}
		\alpha_{11} & \alpha_{12}\\
		\alpha_{21} & \alpha_{22}
	\end{array}\right)^{-1}\left(\begin{array}{c}
		\alpha_{22}+\alpha_{12}\\
		0
	\end{array}\right)\label{eq:p2_ss_matrix}\\*
	& = & \frac{1}{\alpha_{11}\alpha_{22}-\alpha_{12}\alpha_{21}}\left(\begin{array}{cc}
		\alpha_{22} & -\alpha_{12}\\
		-\alpha_{21} & \alpha_{11}
	\end{array}\right)\left(\begin{array}{c}
		\alpha_{22}+\alpha_{12}\\
		0
	\end{array}\right)\nonumber
\end{eqnarray}

\begin{eqnarray*}
	p_{1}^{\infty} & = & \frac{\alpha_{22}\left(\alpha_{22}+\alpha_{12}\right)}{\alpha_{11}\alpha_{22}-\alpha_{12}\alpha_{21}}\\*
	p_{2}^{\infty} & = & \frac{-\alpha_{21}\left(\alpha_{22}+\alpha_{12}\right)}{\alpha_{11}\alpha_{22}-\alpha_{12}\alpha_{21}}
\end{eqnarray*}

\subsection{Linear analysis of the $p_{2}$ model}\label{subsec_linear_analysis_p2_model}

The linear admittance of the $p_{2}$ model can be obtained using a method similar to that previously described for the Hodgkin-Huxley model by \cite{Mauro1970}. The matrix form given in Eq. \ref{eq:p2_ode_matrix} represents the $p_{2}$ model of Eq. \ref{eq:p2_ode} with $k_{1}$, $k_{2}$, $k_{3}$, $k_{4}$ replaced by $\alpha_{11}$, $\alpha_{12}$, $\alpha_{21}$, $\alpha_{22}$ as follows
\begin{eqnarray}
	\frac{dp_{1}}{dt} & = & \alpha_{11}p_{1}+\alpha_{12}p_{2}-\alpha_{22}-\alpha_{12}\label{eq:p2_alphaij}\\*
	\frac{dp_{2}}{dt} & = & \alpha_{21}p_{1}+\alpha_{22}p_{2}\nonumber 
\end{eqnarray}

The linearization for a small perturbation at steady state is given by
\begin{eqnarray*}
	\delta\frac{dp_{1}}{dt} & = & \frac{d\alpha_{11}}{dV}\delta Vp_{1}^{\infty}+\alpha_{11}\delta p_{1}+\frac{d\alpha_{12}}{dV}\delta Vp_{2}^{\infty}+\alpha_{12}\delta p_{2}-\frac{d\left(\alpha_{22}+\alpha_{12}\right)}{dV}\delta V\\*
	\delta\frac{p_{2}}{dt} & = & \frac{d\alpha_{21}}{dV}\delta Vp_{1}^{\infty}+\alpha_{21}\delta p_{1}+\frac{d\alpha_{22}}{dV}\delta Vp_{2}^{\infty}+\alpha_{22}\delta p_{2}
\end{eqnarray*}
or equivalently
\begin{eqnarray*}
	\delta\frac{dp_{1}}{dt}-\alpha_{11}\delta p_{1}-\alpha_{12}\delta p_{2} & = & \left[\frac{d\alpha_{11}}{dV}p_{1}^{\infty}+\frac{d\alpha_{12}}{dV}p_{2}^{\infty}-\frac{d\left(\alpha_{22}+\alpha_{12}\right)}{dV}\right]\delta V\\*
	\delta\frac{p_{2}}{dt}-\alpha_{21}\delta p_{1}-\alpha_{22}\delta p_{2} & = & \left[\frac{d\alpha_{21}}{dV}p_{1}^{\infty}+\frac{d\alpha_{22}}{dV}p_{2}^{\infty}\right]\delta V
\end{eqnarray*}

Applying the Fourier transform with angular frequency $\omega$
\begin{eqnarray*}
	\left(i\omega-\alpha_{11}\right)\widehat{\delta p_{1}}-\alpha_{12}\widehat{\delta p_{2}} & = & \left[\frac{d\alpha_{11}}{dV}p_{1}^{\infty}+\frac{d\alpha_{12}}{dV}p_{2}^{\infty}-\frac{d\left(\alpha_{22}+\alpha_{12}\right)}{dV}\right]\widehat{\delta V}\\*
	\left(i\omega-\alpha_{22}\right)\widehat{\delta p_{2}}-\alpha_{21}\widehat{\delta p_{1}} & = & \left[\frac{d\alpha_{21}}{dV}p_{1}^{\infty}+\frac{d\alpha_{22}}{dV}p_{2}^{\infty}\right]\widehat{\delta V}
\end{eqnarray*}

Then $\widehat{\delta p_{2}}$ can be deduced from
\begin{eqnarray*}
	\alpha_{21}\left(i\omega-\alpha_{11}\right)\widehat{\delta p_{1}}-\alpha_{21}\alpha_{12}\widehat{\delta p_{2}} & = & \alpha_{21}\left[\frac{d\alpha_{11}}{dV}p_{1}^{\infty}+\frac{d\alpha_{12}}{dV}p_{2}^{\infty}-\frac{d\left(\alpha_{22}+\alpha_{12}\right)}{dV}\right]\widehat{\delta V}\\*
	\left(i\omega-\alpha_{11}\right)\left(i\omega-\alpha_{22}\right)\widehat{\delta p_{2}}-\left(i\omega-\alpha_{11}\right)\alpha_{21}\widehat{\delta p_{1}} & = & \left(i\omega-\alpha_{11}\right)\left[\frac{d\alpha_{21}}{dV}p_{1}^{\infty}+\frac{d\alpha_{22}}{dV}p_{2}^{\infty}\right]\widehat{\delta V}
\end{eqnarray*}

By summation
\begin{eqnarray*}
	\left[\left(i\omega-\alpha_{11}\right)\left(i\omega-\alpha_{22}\right)-\alpha_{21}\alpha_{12}\right]\widehat{\delta p_{2}} & = & \alpha_{21}\left[\frac{d\alpha_{11}}{dV}p_{1}^{\infty}+\frac{d\alpha_{12}}{dV}p_{2}^{\infty}-\frac{d\left(\alpha_{22}+\alpha_{12}\right)}{dV}\right]\widehat{\delta V}\\*
	& + & \left(i\omega-\alpha_{11}\right)\left[\frac{d\alpha_{21}}{dV}p_{1}^{\infty}+\frac{d\alpha_{22}}{dV}p_{2}^{\infty}\right]\widehat{\delta V}
\end{eqnarray*}

This implies the rate of variation
\[
\frac{\widehat{\delta p_{2}}}{\widehat{\delta V}}=\frac{\alpha_{21}\left[\frac{d\alpha_{11}}{dV}p_{1}^{\infty}+\frac{d\alpha_{12}}{dV}p_{2}^{\infty}-\frac{d\left(\alpha_{22}+\alpha_{12}\right)}{dV}\right]+\left(i\omega-\alpha_{11}\right)\left[\frac{d\alpha_{21}}{dV}p_{1}^{\infty}+\frac{d\alpha_{22}}{dV}p_{2}^{\infty}\right]}{\left(i\omega-\alpha_{11}\right)\left(i\omega-\alpha_{22}\right)-\alpha_{21}\alpha_{12}}
\]

The current across the lipid bilayer is given by
\[
I=C_{m}\frac{dV}{dt}+g_{\mathrm{L}}\left(V-V_{\mathrm{L}}\right)+g_{\mathrm{K}}p_{2}\left(V-V_{\mathrm{K}}\right)
\]

It is linearized by
\[
\delta I=C_{m}\frac{d\delta V}{dt}+g_{\mathrm{L}}\delta V+g_{\mathrm{K}}\left[\delta p_{2}\left(V_{0}-V_{\mathrm{K}}\right)+p_{2}^{\infty}\delta V\right]
\]

Applying the Fourier transform with angular frequency $\omega$
\[
\widehat{\delta I}=C_{m}i\omega\widehat{\delta V}+g_{\mathrm{L}}\widehat{\delta V}+g_{\mathrm{K}}\left[\widehat{\delta p_{2}}\left(V_{0}-V_{\mathrm{K}}\right)+p_{2}^{\infty}\widehat{\delta V}\right]
\]

This provides the admittance of the $p_{2}$ model
\[
\widehat{Y}=\frac{\widehat{\delta I}}{\widehat{\delta V}}=C_{m}i\omega+g_{\mathrm{L}}+g_{\mathrm{K}}\left[\frac{\widehat{\delta p_{2}}}{\widehat{\delta V}}\left(V_{0}-V_{\mathrm{K}}\right)+p_{2}^{\infty}\right]
\]

\subsection{Markov power spectra of the $p_{2}$ model}\label{subsec_markov_power_spectra_p2_model}

From Eq. \ref{eq:p2_ode_matrix} and Eq. \ref{eq:p2_ss_matrix}, the differential equation can be written in the homogeneous form
\[
\left(\begin{array}{c}
	\frac{d\left(p_{1}-p_{1}^{\infty}\right)}{dt}\\
	\frac{d\left(p_{2}-p_{2}^{\infty}\right)}{dt}
\end{array}\right)=\left(\begin{array}{cc}
	\alpha_{11} & \alpha_{12}\\
	\alpha_{21} & \alpha_{22}
\end{array}\right)\left(\begin{array}{c}
	p_{1}-p_{1}^{\infty}\\
	p_{2}-p_{2}^{\infty}
\end{array}\right)
\]

More concisely
\[
\frac{d\mathbf{p}}{dt}=\mathbf{M}\mathbf{p}
\]
where $\mathbf{p}=\left(\begin{array}{c}
	p_{1}-p_{1}^{\infty}\\
	p_{2}-p_{2}^{\infty}
\end{array}\right)$ and $\mathbf{M}=\left(\begin{array}{cc}
	\alpha_{11} & \alpha_{12}\\
	\alpha_{21} & \alpha_{22}
\end{array}\right)$.

The general solution is a linear combination
\begin{equation}
	\mathbf{p}=c_{1}e^{\lambda_{1}t}\mathbf{v_{1}}+c_{2}e^{\lambda_{2}t}\mathbf{v_{2}}\label{eq:p2_general_solution}
\end{equation}
where $\lambda_{1}$, $\lambda_{2}$ are eigenvalues of $\mathbf{M}$, $\mathbf{v_{1}}$, $\mathbf{v_{2}}$ are eigenvectors of $\mathbf{M}$ and $c_{1}$, $c_{2}$ are constants. Indeed
\begin{eqnarray*}
	\mathbf{M}\mathbf{p} & = & c_{1}e^{\lambda_{1}t}\mathbf{M}\mathbf{v_{1}}+c_{2}e^{\lambda_{2}t}\mathbf{M}\mathbf{v_{2}}\\
	& = & c_{1}e^{\lambda_{1}t}\lambda_{1}\mathbf{v_{1}}+c_{2}e^{\lambda_{2}t}\lambda_{2}\mathbf{v_{2}}\\
	& = & \frac{d\mathbf{p}}{dt}
\end{eqnarray*}

The trace $T=\alpha_{11}+\alpha_{22}$ and determinant $D=\alpha_{11}\alpha_{22}-\alpha_{12}\alpha_{21}$ of $\mathbf{M}$ determine the eigenvalues
\[
\lambda_{1,2}=\frac{T\pm\sqrt{T^{2}-4D}}{2}
\]
\[
\lambda_{1,2}=\frac{\alpha_{11}+\alpha_{22}\pm\sqrt{\left(\alpha_{11}+\alpha_{22}\right)^{2}-4\left(\alpha_{11}\alpha_{22}-\alpha_{12}\alpha_{21}\right)}}{2}
\]

Since $\mathbf{p}$ represents probabilities, $\lambda_{1}$ and $\lambda_{2}$ must be negative to have exponential decreases rather than exponential increases. This determines two time constants for the $p_{2}$ model
\[
\tau_{1}=-\frac{1}{\lambda_{1}}\textrm{ and }\tau_{2}=-\frac{1}{\lambda_{2}}
\]

The eigenvectors $\mathbf{v_{1}}$ and $\mathbf{v_{2}}$ are obtained by solving the equation $\mathbf{M}\mathbf{v}=\lambda\mathbf{v}$ and substituting eigenvalues $\lambda=\lambda_{1}$ and $\lambda=\lambda_{2}$
\[
\mathbf{v_{1}}=\left(\begin{array}{c}
	\frac{\lambda_{1}-\alpha_{22}}{\alpha_{21}}\\
	1
\end{array}\right)
\]
\[
\mathbf{v_{2}}=\left(\begin{array}{c}
	\frac{\lambda_{2}-\alpha_{22}}{\alpha_{21}}\\
	1
\end{array}\right)
\]

Inserting $\mathbf{v_{1}}$ and $\mathbf{v_{2}}$ into Eq. \ref{eq:p2_general_solution}
\[
\left(\begin{array}{c}
	p_{1}-p_{1}^{\infty}\\
	p_{2}-p_{2}^{\infty}
\end{array}\right)=c_{1}e^{\lambda_{1}t}\left(\begin{array}{c}
	\frac{\lambda_{1}-\alpha_{22}}{\alpha_{21}}\\
	1
\end{array}\right)+c_{2}e^{\lambda_{2}t}\left(\begin{array}{c}
	\frac{\lambda_{2}-\alpha_{22}}{\alpha_{21}}\\
	1
\end{array}\right)
\]

The time $t=0$ determines the coefficients $c_{1}$ and $c_{2}$ as follows
\begin{eqnarray*}
	p_{1}\left(0\right)-p_{1}^{\infty} & = & c_{1}\frac{\lambda_{1}-\alpha_{22}}{\alpha_{21}}+c_{2}\frac{\lambda_{2}-\alpha_{22}}{\alpha_{21}}\\*
	p_{2}\left(0\right)-p_{2}^{\infty} & = & c_{1}+c_{2}
\end{eqnarray*}
\begin{eqnarray*}
	p_{1}\left(0\right)-p_{1}^{\infty} & = & c_{1}\frac{\lambda_{1}-\alpha_{22}}{\alpha_{21}}+\left(p_{2}\left(0\right)-p_{2}^{\infty}-c_{1}\right)\frac{\lambda_{2}-\alpha_{22}}{\alpha_{21}}\\*
	c_{2} & = & p_{2}\left(0\right)-p_{2}^{\infty}-c_{1}
\end{eqnarray*}
\begin{eqnarray*}
	c_{1}\left(\frac{\lambda_{2}-\lambda_{1}}{\alpha_{21}}\right) & = & p_{2}\left(0\right)\frac{\lambda_{2}-\alpha_{22}}{\alpha_{21}}-p_{2}^{\infty}\frac{\lambda_{2}-\alpha_{22}}{\alpha_{21}}-p_{1}\left(0\right)+p_{1}^{\infty}\\*
	c_{2} & = & p_{2}\left(0\right)-p_{2}^{\infty}-c_{1}
\end{eqnarray*}
\begin{eqnarray*}
	c_{1} & = & \frac{\left(\lambda_{2}-\alpha_{22}\right)\left(p_{2}\left(0\right)-p_{2}^{\infty}\right)-\alpha_{21}\left(p_{1}\left(0\right)-p_{1}^{\infty}\right)}{\lambda_{2}-\lambda_{1}}\\*
	c_{2} & = & p_{2}\left(0\right)-p_{2}^{\infty}-c_{1}
\end{eqnarray*}

The power spectrum can be calculated using the same reasoning as with the model $n^{4}$ previously because it is a sequential model with $3$ states instead of $5$ states. In particular, the single-channel autocovariance is based on the conditional probability $p_{2}\left(t\vert0\right)$ and the steady state probability $p_{2}^{\infty}$

\[
C\left(t\right)=p_{2}^{\infty}p_{2}\left(t\vert0\right)-\left(p_{2}^{\infty}\right)^{2}
\]

Considering the unique state $\left(p_{0},p_{1},p_{2}\right)=\left(0,0,1\right)$ when the channel is open
\begin{eqnarray*}
	c_{1} & = & \frac{\left(\lambda_{2}-\alpha_{22}\right)\left(1-p_{2}^{\infty}\right)+\alpha_{21}p_{1}^{\infty}}{\lambda_{2}-\lambda_{1}}\\*
	c_{2} & = & 1-p_{2}^{\infty}-c_{1}
\end{eqnarray*}

As a result, the conditional probability $p_{2}\left(t\vert0\right)$ is given by
\[
p_{2}\left(t\vert0\right)=p_{2}^{\infty}+c_{1}e^{\lambda_{1}t}+c_{2}e^{\lambda_{2}t}
\]

The autocovariance is given by
\begin{eqnarray*}
	C\left(t\right) & = & p_{2}^{\infty}\left[p_{2}^{\infty}+c_{1}e^{\lambda_{1}t}+c_{2}e^{\lambda_{2}t}\right]-\left(p_{2}^{\infty}\right)^{2}\\*
	& = & p_{2}^{\infty}c_{1}e^{\lambda_{1}t}+p_{2}^{\infty}c_{2}e^{\lambda_{2}t}
\end{eqnarray*}

This implies autocovariance $C_{I_{\mathrm{K}}}\left(t\right)$ of the total fluctuating current
\begin{eqnarray*}
	C_{I_{\mathrm{K}}}\left(t\right) & = & N_{\mathrm{K}}i_{\mathrm{K}}^{2}C\left(t\right)\\*
	& = & N_{\mathrm{K}}i_{\mathrm{K}}^{2}p_{2}^{\infty}\left(c_{1}e^{\lambda_{1}t}+c_{2}e^{\lambda_{2}t}\right)
\end{eqnarray*}

Following the same conventions as previously, the power spectrum is
\begin{eqnarray*}
	S_{I_{\mathrm{K}}}\left(\omega\right) & = & 2S_{I_{\mathrm{K}}}^{\pm}\left(\omega\right)\\*
	& = & 4\Re\int_{0}^{+\infty}C_{I_{\mathrm{K}}}\left(t\right)e^{-i\omega t}dt\\*
	& = & 4N_{\mathrm{K}}i_{\mathrm{K}}^{2}p_{2}^{\infty}\Re\int_{0}^{+\infty}\left(c_{1}e^{\lambda_{1}t}+c_{2}e^{\lambda_{2}t}\right)e^{-i\omega t}dt
\end{eqnarray*}

There are two integrals to be calculated for each number $q=1,2$
\[
S_{q}\left(\omega\right)=\int_{0}^{+\infty}c_{q}e^{\lambda_{q}t}e^{-i\omega t}dt
\]

Using the non-unitary Fourier transform $\mathcal{F}\left[e^{-a\left|t\right|}\right]=\frac{2a}{a^{2}+\omega^{2}}$ with angular frequencies and divided by $2$ because of the positive half of the time $\left[0;+\infty\right]$
\[
S_{q}\left(\omega\right)=c_{q}\frac{-\lambda_{q}}{\lambda_{q}^{2}+\omega^{2}}
\]

Then, the power spectrum is deduced
\[
S_{I_{\mathrm{K}}}\left(\omega\right)=4N_{\mathrm{K}}i_{\mathrm{K}}^{2}p_{2}^{\infty}\left(c_{1}\frac{-\lambda_{1}}{\lambda_{1}^{2}+\omega^{2}}+c_{q}\frac{-\lambda_{2}}{\lambda_{2}^{2}+\omega^{2}}\right)
\]

By using the time constants $\tau_{1}=-\frac{1}{\lambda_{1}}$ and $\tau_{2}=-\frac{1}{\lambda_{2}}$
\begin{eqnarray*}
	S_{I_{\mathrm{K}}}\left(\omega\right) & = & 4N_{\mathrm{K}}i_{\mathrm{K}}^{2}p_{2}^{\infty}\left(c_{1}\frac{1/\tau_{1}}{1/\tau_{1}^{2}+\omega^{2}}+c_{q}\frac{1/\tau_{2}}{1/\tau_{2}^{2}+\omega^{2}}\right)\\*
	& = & 4N_{\mathrm{K}}i_{\mathrm{K}}^{2}p_{2}^{\infty}\left(c_{1}\frac{\tau_{1}}{1+\omega^{2}\tau_{1}^{2}}+c_{q}\frac{\tau_{2}}{1+\omega^{2}\tau_{2}^{2}}\right)\label{eq:rossum_power_spectrum_p2}
\end{eqnarray*}

\begin{figure}[H]
\centering
\includegraphics[width=12cm]{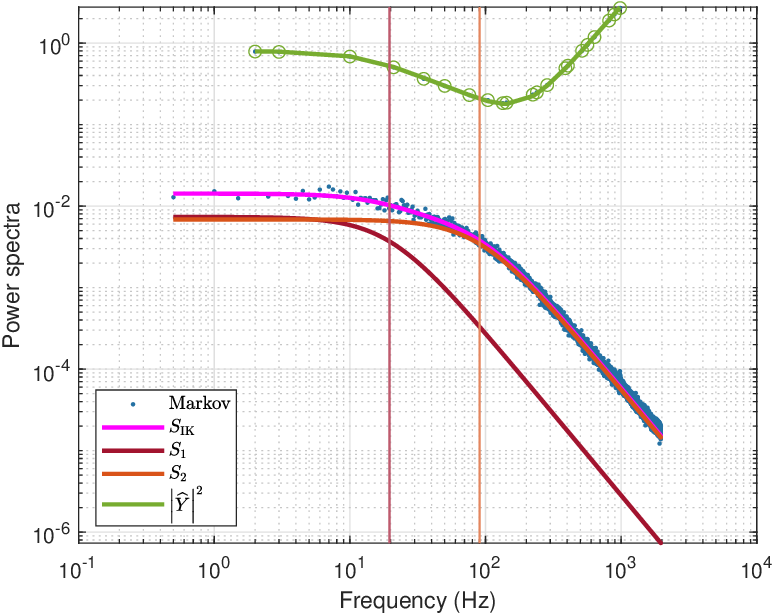}
\caption{\label{fig:qsa_vc_p2_5mV_M128_noise_stimulated}\textbf{Markov simulation of the $p_{2}$ model for a $5\mathrm{mV}$ depolarization with stimulus amplitude $0.25\mathrm{mV}$.} The power spectra approximated by $128$ iterations are represented by the blue scatterplot. The predicted power spectrum $S_{I_{\mathrm{K}}}$ (magenta curve) accurately fits the blue scatterplot. The two Lorentzians compose the predicted power spectrum with $S_{1}$ (brown curve) and $S_{2}$ (orange curve). The corresponding corner frequencies $\omega_{1}$ and $\omega_{2}$ are represented by vertical lines. The squared admittance $\left|\widehat{Y}\right|^{2}$ (green curve) matches the linear Markov responses after adjustment of the vertical offset. The QSA multi-sinusoidal stimulus frequencies are {[}2, 3, 10, 21, 35, 50, 76, 104, 134, 143, 223, 239, 285, 388,405, 515, 564, 636, 815, 892, 982{]} Hertz.}
\end{figure}

\begin{figure}[H]
\centering
\includegraphics[width=12cm]{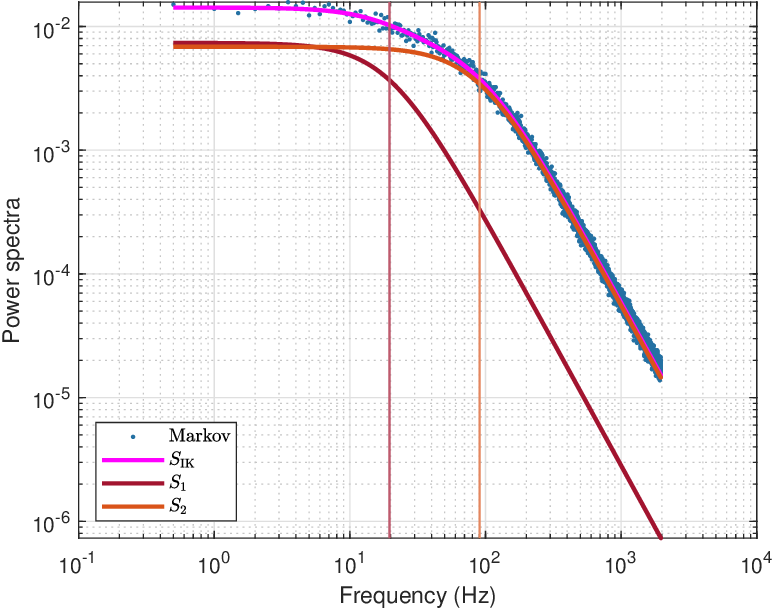}
\caption{\label{fig:qsa_vc_p2_5mV_M128_noise_notstimulated}\textbf{Markov simulation of the $p_{2}$ model for a $5\mathrm{mV}$ depolarization without stimulus.} The fluctuation power spectra are the same as those in Fig. \ref{fig:qsa_vc_p2_5mV_M128_noise_stimulated} but there is no admittance.}
\end{figure}

Fig. \ref{fig:qsa_vc_p2_5mV_M128_noise_stimulated} illustrates a voltage clamp Markov simulation for a $5\mathrm{mV}$ depolarization of the spontaneous potassium current fluctuations superimposed on the theoretically predicted power spectrum $S_{\mathrm{IK}}$, the two Lorentzians, the two corner frequencies and the squared admittance $\left|\widehat{Y}\right|^{2}$. The simulation was done with the Markov model and QSA multi-sinusoidal stimuli of amplitude $0.25\mathrm{mV}$. At this potential, the predicted power spectrum $S_{\mathrm{IK}}$ (magenta curve) is well described by a single Lorentzian function for frequencies greater than the corner frequency. The square of the admittance is obtained with an input of constant amplitude, which makes it possible to compare it to other curves independent of the input. Fig. \ref{fig:qsa_vc_p2_5mV_M128_noise_notstimulated} shows that simulations done without stimulus gave identical power spectra.

\subsection{Neuronal functions compared to neuronal fluctuations}\label{subsec_p2_functions_vs_fluctuations}

Individual neurons are characterized both by deterministic neuronal functions, such as those revealed by linear and quadratic analysis (QSA), and by random fluctuations, such as those produced by Markov models. Both behaviors play an essential role in subthreshold neuronal processing to generate action potentials and, consequently, for neuronal networks in general.

Both behaviors are fundamentally related to the nonlinear kinetic processes underlying ion channels, such as the $n^{4}$ model, its $n^{2}$ simplification, or the generalization of $n^{2}$ to the $p_{2}$ model without exponentiation, as described above. In contrast, the fluctuation-dissipation theorem implies that ion channel fluctuations should be described by a linearization of the underlying nonlinear channel kinetics. To address these issues, the sequential kinetic model $p_{2}$ described above is an approximate alternative to the Hodgkin-Huxley model that may be useful because it is not based on exponentiation. In all cases, Markov models are considered an adequate approach to simulate the fluctuations of a single-channel. Other neuronal noise fluctuation models are of interest, such as stochastic differential equations (SDEs), but are not considered in this paper. Since sequential kinetic models include Hodgkin-Huxley models with exponentiation as a special case, and have been shown to describe kinetic behavior as well as or better than the Hodgkin-Huxley model, it is appropriate to compare the simple $p_{2}$ model with the $n^{4}$ model. This will be done in the following voltage clamp simulations for the model with nonlinear exponentiation $n^{4}$ and the model without exponentiation $p_{2}$ to provide a comparison of Markov fluctuations, their linear and quadratic (QSA) stimulated behavior. A fundamental question is whether models with exponentiation of the Hodgkin-Huxley type or models without exponentiation are the best descriptions of a single neuron behavior. These fluctuation simulations determine whether or not the spontaneous current noise at a fixed voltage clamped membrane potential should be based on a nonlinear exponentiation such that $n^{4}$.

The control case for nonlinear exponentiation is the Hodgkin-Huxley $n^{4}$ model. The frequency domain analysis is performed on simulated data measurements at two different depolarized membrane potentials, namely $5\mathrm{mV}$ and $55\mathrm{mV}$, such as one could record from a voltage clamped biological neuron. These two voltage clamp potentials were chosen to be either near the resting value or full activation of the voltage-dependent ionic conductances. Simulations were performed using MATLAB R2022b (MathWorks) with the nonlinear classical Hodgkin-Huxley $n^{4}$ model described in above sections. Two modes have been considered, deterministic and stochastic, based on the programs of \cite{Goldwyn2011a,Goldwyn2011b}. The deterministic mode was done using Euler's method for ODE, while the stochastic mode was based on Gillespie's algorithm. The results of the numerical simulations are superimposed on an analytical form for the impedance (Eq. \ref{eq:mauro_hh_impedance}) and the Markov fluctuations (Eq. \ref{eq:rossum_power_spectrum_n4}). Thus, the simulations are analogous to measured data from a biological neuron, which can be analyzed in the frequency domain using the methods described above. As demonstrated by \cite{Magnani2013}, this type of frequency domain analysis is an especially efficient and useful way to determine the accuracy of a particular model, since frequency domain analysis of data from biological neurons is a sensitive experimental measure that can be rigorously compared to model predictions. When data are produced by a model, as in this paper, comparisons between numerical simulations and analytical expressions of impedance and Markov fluctuations are used to check the accuracy of calculations. In addition, the QSA method provides a nonlinear analysis independent of a particular type of model or experiment, and thus can be applied to compare simulated data for different models.

\begin{figure}[H]
\centering
\includegraphics[width=14cm]{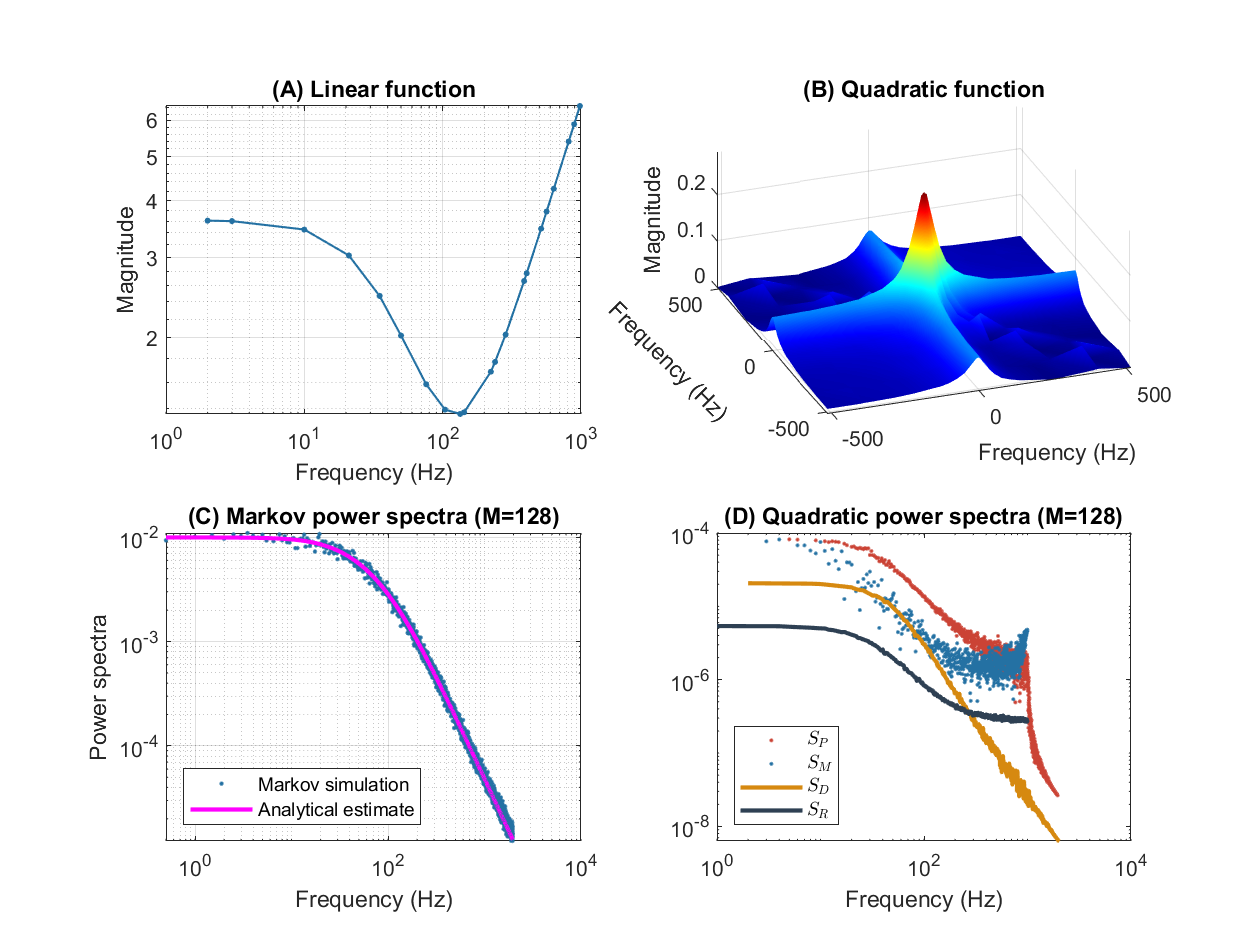}
\caption{\label{fig:qsa_vc_n4_5mV_M128}\textbf{Frequency analysis of the $n^{4}$ model for a $5\mathrm{mV}$ depolarization.} \textbf{(A)} Linear admittance (smooth line) generated by a QSA multi-sinusoidal stimulus of amplitude $0.25\mathrm{mV}$ and frequencies {[}2, 3, 10, 21, 35, 50, 76, 104, 134, 143, 223, 239, 285, 388,405, 515, 564, 636, 815, 892, 982{]} Hertz. The linear responses at stimulating frequencies are marked by small circles. \textbf{(B)} QSA matrix in 3D representation obtained with the same stimulus as the linear function, but the plot is cut at $500\mathrm{Hz}$ for a better readability. The color code indicates the amplitude of the quadratic response. Each value $\left(\omega_{i},\omega_{j},\left|Q_{i,j}\right|\right)$ in the 3D plot represents the magnitude of the quadratic response to a frequency interaction. \textbf{(C)} Power spectra of the Markov simulation ($128$ iterations) superimposed on the analytical estimate. The frequencies are continuous up to twice the maximum stimulus frequency to include the highest QSA frequency (frequency doubling). The analytical estimation provides an excellent control of the Markov simulation. \textbf{(D)} Quadratic power spectra averaged over several ODE simulations ($128$ iterations) for different sets of QSA frequencies randomized up to $1000\mathrm{Hz}$. Quadratic responses are represented at their particular frequencies, namely $S_{P}\left(\left|\omega_{i}\right|+\left|\omega_{j}\right|\right)$ at frequency sums (red points), $S_{M}\left(\left|\left|\omega_{i}\right|-\left|\omega_{j}\right|\right|\right)$ at frequency differences (blue points), $S_{D}\left(2\left|\omega_{k}\right|\right)$ at frequency doubling (orange points) and $S_{R}\left(\left|\omega_{j}\right|\right)$ for the mean squared quadratic output by matrix columns (black points). Unlike the QSA matrix which is based on ratios of outputs to inputs, quadratic power spectra are evaluated directly from output measurements like Markov power spectra.}
\end{figure}

Fig. \ref{fig:qsa_vc_n4_5mV_M128} illustrates the frequency analysis of the $n^{4}$ model for a $5\mathrm{mV}$ depolarization. The upper left plot (Fig. \ref{fig:qsa_vc_n4_5mV_M128}A) shows a typical low frequency linear admittance anti-resonance (smooth line) generated by a QSA multi-sinusoidal stimulus of amplitude $0.25\mathrm{mV}$. The upper right plot (Fig. \ref{fig:qsa_vc_n4_5mV_M128}B) shows the QSA matrix as a 3D representation, an intersection of lines on the plane for any two linear frequencies $\omega_{i}$, $\omega_{j}$ represents an interactive quadratic frequency $\omega_{i}+\omega_{j}$ for $i,j\in\left\{ -N,\ldots,-1,+1,\ldots,+N\right\} $ and the color code indicates the amplitude of the quadratic response. Remarkably, the quadratic response shows no anti-resonance. The lower left plot (Fig. \ref{fig:qsa_vc_n4_5mV_M128}C) shows the power spectra of the Markov simulation superimposed on the analytical estimate of the Eq. \ref{eq:rossum_power_spectrum_n4}. It is clear that the analytical estimation provides an excellent control of the Markov simulation, both of which reveal the characteristics of a low-pass filter. The lower right plot (Fig. \ref{fig:qsa_vc_n4_5mV_M128}D) shows the quadratic power spectra averaged over several ODE simulations for different random sets of QSA frequencies. Indeed, since the QSA frequency sets have few frequencies, it is necessary to average several measurements to increase the accuracy of the power spectra. Individual amplitudes of the various quadratic responses are represented at their particular frequencies, namely $S_{P}\left(\left|\omega_{i}\right|+\left|\omega_{j}\right|\right)$ at frequency sums (red points), $S_{M}\left(\left|\left|\omega_{i}\right|-\left|\omega_{j}\right|\right|\right)$ at frequency differences (blue points), $S_{D}\left(2\left|\omega_{k}\right|\right)$ at frequency doubling (orange points) and $S_{R}\left(\left|\omega_{j}\right|\right)$ for the mean squared quadratic output by matrix columns (black points). All of these responses exhibit low-pass filter characteristics, but they flatten out by reaching a constant value at high frequencies, with the exception of $S_{D}\left(2\left|\omega_{k}\right|\right)$ at frequency doubling for which the amplitudes decrease at high frequencies. Unlike the QSA matrix which is based on ratios of outputs to inputs, quadratic power spectra are evaluated directly from output measurements like Markov power spectra.

The amplitudes of the quadratic power spectra are smaller than those of the Markov power spectra. Indeed, quadratic responses are an order of magnitude smaller than linear responses, while the fluctuation-dissipation theorem relates spontaneous fluctuations to linear responses. In particular, Markov power spectra are based on autocorrelation, which corresponds to the second order cumulant, whereas higher order spectra would involve at least the third order cumulant as explained by \cite{Mendel1991}. At this end of this paper, the discussion provides some simulations to compare the amplitudes of quadratic responses and spontaneous fluctuations.

It is well known that nonlinear systems can generate frequency mixing processes, such as those producing interactive frequencies $\left|\omega_{i}\right|+\left|\omega_{j}\right|$, $\left|\left|\omega_{i}\right|-\left|\omega_{j}\right|\right|$ and $2\left|\omega_{k}\right|$. Frequency mixing behavior is fundamental in some scientific fields, such as nonlinear optics, as described for example by \cite{Boyd2008}. Frequency mixing introduces complexity into the response, which may contain frequencies that are not present in the stimulus. Thus, neurons mix stimulus oscillations into quadratic responses that may appear less uniform than linear responses. In particular, the power spectra $S_{P}\left(\left|\omega_{i}\right|+\left|\omega_{j}\right|\right)$ and $S_{M}\left(\left|\left|\omega_{i}\right|-\left|\omega_{j}\right|\right|\right)$ show scatter plots with a lot of dispersion that look like stochastic fluctuations. It should be noted that although different random sets of QSA frequencies were used for averaging, each individual QSA matrix is obtained from a deterministic ODE. Randomizing the frequencies only extends the range of frequencies for analysis, implying that the dispersion is due to the quadratic neuronal function rather than a lack of frequencies. In contrast, frequency doubling generates power spectra $S_{D}\left(2\left|\omega_{k}\right|\right)$ that approximately follow a smooth line. This is because frequency doubling is calculated along the frequency diagonal $\left(\omega_{k},\omega_{k}\right)$ on the 3D representation, which decreases smoothly from low frequencies (color-coded red for height) to high frequencies (color-coded blue for height). Similarly, the power spectra $S_{R}\left(\left|\omega_{j}\right|\right)$ for the mean squared quadratic output by matrix columns appear to be quite smooth for the reason that the 3D representation is smooth and summed column by column. Remarkably, the power spectra $S_{D}\left(2\left|\omega_{k}\right|\right)$ tends to zero at high frequencies due to the attenuation of quadratic responses along the diagonal $\left(\omega_{k},\omega_{k}\right)$, while the power spectra $S_{R}\left(\left|\omega_{j}\right|\right)$ flatten at high frequencies because of the residual quadratic responses around the horizontal $\left(x,0\right)$ and vertical axes $\left(0,y\right)$. Therefore, the irregularity of the power spectra $S_{P}\left(\left|\omega_{i}\right|+\left|\omega_{j}\right|\right)$ and $S_{M}\left(\left|\left|\omega_{i}\right|-\left|\omega_{j}\right|\right|\right)$ are fundamentally due to non-trivial quadratic frequency interactions between $\omega_{i}$ and $\omega_{j}$ when they do not follow the smooth shape of the 3D representation (by diagonal or by columns).

\begin{figure}[H]
\centering
\includegraphics[width=12cm]{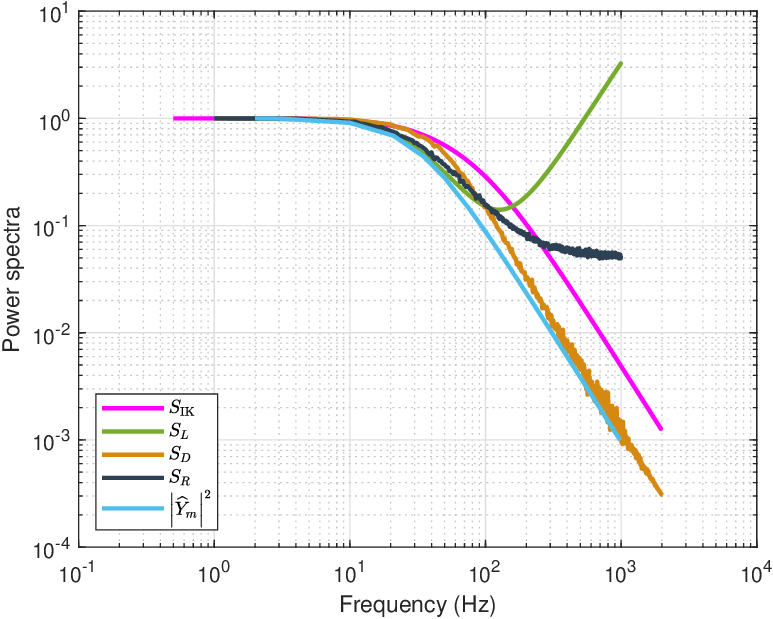}
\caption{\label{fig:qsa_vc_n4_5mV_M128_comparison}\textbf{Power spectral analysis of the $n^{4}$ model for a $5\mathrm{mV}$ depolarization.} Comparison between the analytical estimate of Markov fluctuations $S_{\mathrm{IK}}$ (magenta), the linear power spectrum $S_{L}$ (green), the quadratic power spectra $S_{D}$ (orange) and $S_{R}$ (black), the modified squared admittance $\left|\widehat{Y}_{m}\right|^{2}$ (blue).}
\end{figure}

Fig. \ref{fig:qsa_vc_n4_5mV_M128_comparison} shows the superposition of five curves of the $n^{4}$ model for a $5\mathrm{mV}$ depolarization. This allows a comparison of the linear, nonlinear and fluctuation amplitudes behavior. The curves were scaled to unity at low frequencies. The curve $S_{\mathrm{IK}}$ represents the analytical estimate of Markov fluctuations. The curve $S_{L}$ represents the linear power spectrum at fundamental frequencies. The curve $S_{D}$ represents the quadratic power spectrum at frequency doubling. The curve $S_{R}$ represents the mean squared quadratic output by matrix columns. The multi-sinusoidal power spectra $S_{L}$, $S_{D}$, $S_{R}$ were averaged over different random sets of non-overlapping frequencies as in Fig. \ref{fig:qsa_vc_n4_5mV_M128}. The curve $\left|\widehat{Y}_{m}\right|^{2}$ represents the squared admittance modified from the Eq. \ref{eq:mauro_hh_admittance} without the membrane capacitance nor the frequency independent terms, namely
\[
\widehat{Y}_{m}=g_{\mathrm{K}}\left[4n_{0}^{3}\left(V_{0}-V_{\mathrm{K}}\right)\widehat{D_{n}}\right]
\]

The contribution of the capacitance to the actual admittance is in part responsible for an anti-resonance that leads to high frequency responses that are dominant, thus preventing a direct comparison of the linearized admittance with nonlinear responses or fluctuation behavior. The square of the admittance is obtained with an input of constant amplitude, which makes it possible to compare it to other curves independent of the input.

The fall of the modified squared admittance $\left|\widehat{Y}_{m}\right|^{2}$ is more marked than that of the Markov fluctuations $S_{\mathrm{IK}}$, which is also the case for the linear power spectrum $S_{L}$ over a limited range of frequencies before the reversal. The fall of the quadratic power spectra $S_{D}$ and $S_{R}$ is more marked than that of the Markov fluctuations $S_{\mathrm{IK}}$ but less than that of the modified squared admittance, as well as compared to the linear power spectrum $S_{L}$ before the reversal. As explained previously, $S_{D}$ and $S_{R}$ follow the smooth 3D representation of the QSA matrices, in particular $S_{D}$ decreases indefinitely while $S_{R}$ flattens at high frequencies.

These results show that the evoked nonlinear responses have a lower frequency behavior than spontaneous fluctuations, suggesting that Markov fluctuations have a different origin than the nonlinearity evoked by stimulus despite nonlinearity of rate constants or exponentiation $n^{4}$ in the stochastic equations.

\begin{figure}[H]
\centering
\includegraphics[width=14cm]{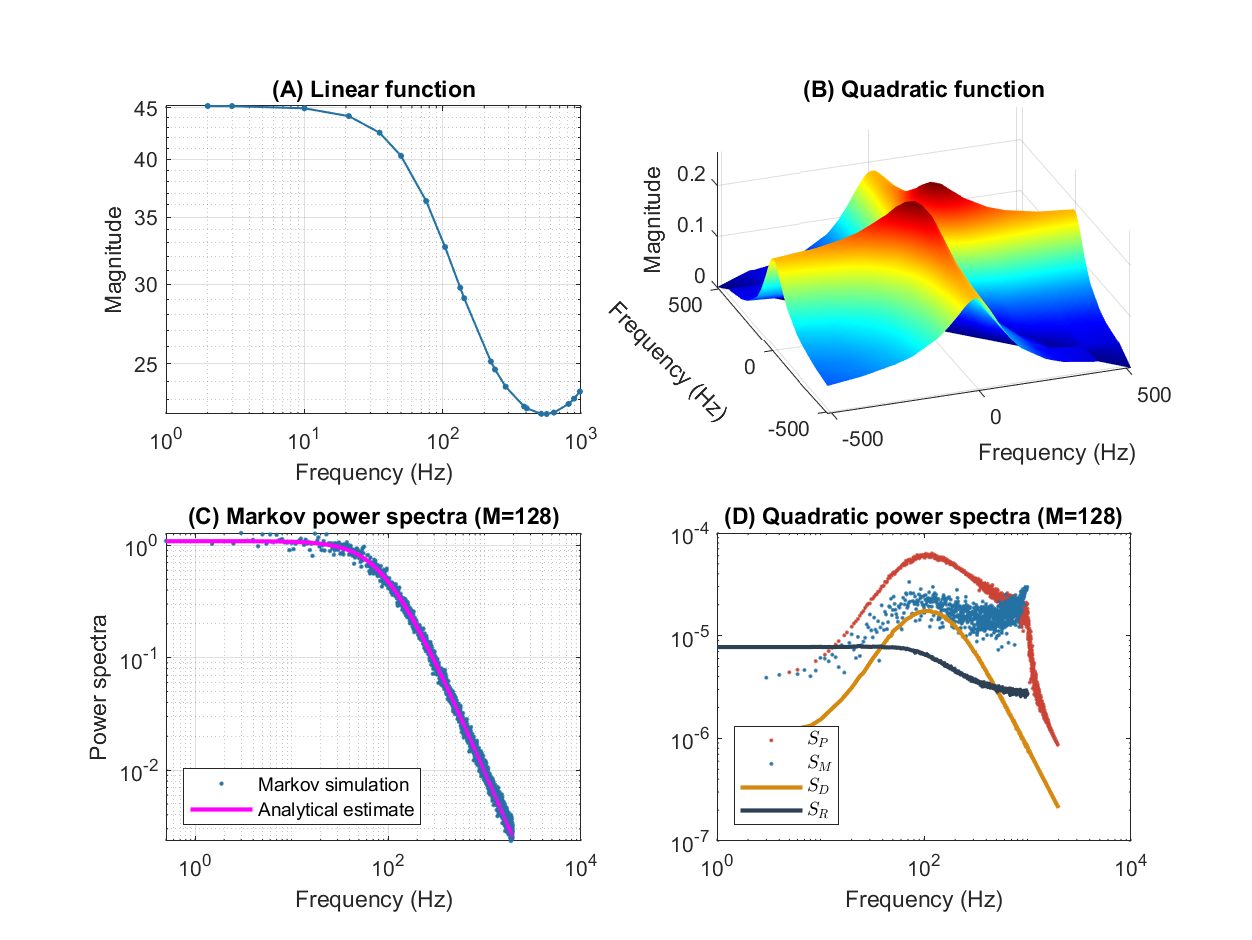}
\caption{\label{fig:qsa_vc_n4_55mV_M128}\textbf{Frequency analysis of the $n^{4}$ model for a $55\mathrm{mV}$ depolarization.} The plots were generated using the same presentation as in Fig. \ref{fig:qsa_vc_n4_5mV_M128}.}
\end{figure}

Fig. \ref{fig:qsa_vc_n4_55mV_M128} shows the frequency analysis of the $n^{4}$ model for a $55\mathrm{mV}$ depolarization, which is very different from the previous $5\mathrm{mV}$ depolarization, clearly illustrating the voltage dependence of the potassium channel in both the linear and nonlinear analyses. The upper left plot shows linear admittance (smooth line) generated by a QSA multi-sinusoidal stimulus of amplitude $0.25\mathrm{mV}$, which reveals the onset of a high frequency anti-resonance minimum. The upper right plot shows the QSA matrix as a 3D representation, which is increased in a slightly shifted bandwidth from the previous depolarization of $5\mathrm{mV}$. The lower left plot shows the power spectra of the Markov simulation superimposed on the analytical estimate, which is similar to a low-pass filter as for a depolarization of $5\mathrm{mV}$. The lower right plot shows the quadratic power spectra averaged over several ODE simulations for different random sets of QSA frequencies. The individual components $S_{P}$, $S_{M}$, $S_{D}$ confirms the slightly shifted bandwidth observed on the QSA matrix, revealing more precisely the marked resonance. In particular, the nonlinear resonant frequencies are clearly lower than that of the linear anti-resonance. However, the $S_{R}$ component based on each matrix column has low-pass filter characteristics. Thus, low-pass behavior is observed at $5\mathrm{mV}$ and $55\mathrm{mV}$ for the Markov simulation of and the column-mean-square $S_{R}$. Interestingly, the nonlinear behavior for highly activated conductances shows resonance in the quadratic responses at frequencies different from the anti-resonance observed for the linear response.

\begin{figure}[H]
\centering
\includegraphics[width=12cm]{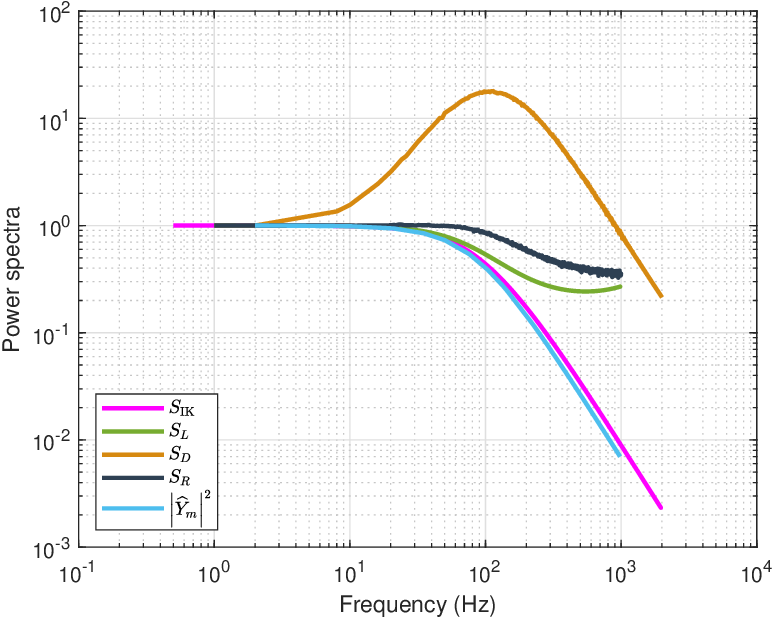}
\caption{\label{fig:qsa_vc_n4_55mV_M128_comparison}\textbf{Power spectral analysis of the $n^{4}$ model for a $55\mathrm{mV}$ depolarization.} The plots were generated using the same presentation as in Fig. \ref{fig:qsa_vc_n4_5mV_M128_comparison}.}
\end{figure}

Fig. \ref{fig:qsa_vc_n4_55mV_M128_comparison} shows the superposition of five curves of the $n^{4}$ model for a $55\mathrm{mV}$ depolarization, using the same presentation as in Fig. \ref{fig:qsa_vc_n4_5mV_M128_comparison}. The modified squared admittance $\left|\widehat{Y}_{m}\right|^{2}$ nearly superimposes on the Markov fluctuations $S_{\mathrm{IK}}$, which is an effect of the voltage-dependent potassium conductance induced by the $55\mathrm{mV}$ depolarization.

\begin{figure}[H]
\centering
\includegraphics[width=12cm]{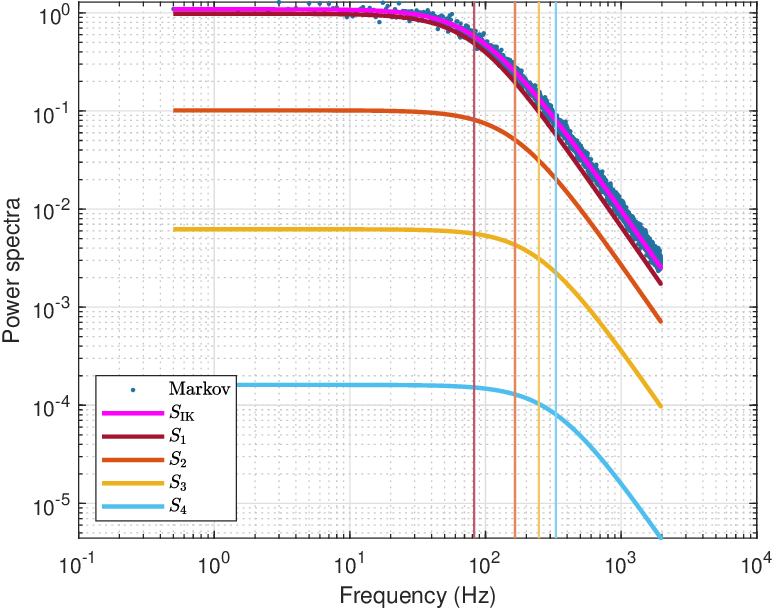}
\caption{\label{fig:qsa_vc_n4_55mV_M128_noise_notstimulated}\textbf{Markov simulation of the $n^{4}$ model for a $55\mathrm{mV}$ depolarization without stimulus.} The plots were generated using the same presentation as in Fig. \ref{fig:qsa_vc_n4_5mV_M128_noise_notstimulated}.}
\end{figure}

Also, by comparing Fig. \ref{fig:qsa_vc_n4_5mV_M128_noise_notstimulated} ($5\mathrm{mV}$) and Fig. \ref{fig:qsa_vc_n4_55mV_M128_noise_notstimulated} ($55\mathrm{mV}$), each of the four relaxation time constants clearly appears as a function of the membrane potential and at large depolarizations the slowest time constant is dominant, i.e. the component $S_{1}$ fits the Markov fluctuations $S_{\mathrm{IK}}$ accurately.

Thus, at $55\mathrm{mV}$ depolarization, the fluctuation behavior $S_{\mathrm{IK}}$ is dominated by $S_{1}$ and has a kinetic behavior similar to that of a modified squared admittance $\left|\widehat{Y}_{m}\right|^{2}$, which does not occur at the less depolarized potential $5\mathrm{mV}$ shown above.

The linear power spectrum $S_{L}$ is similar to the modified squared admittance $\left|\widehat{Y}_{m}\right|^{2}$ at low frequencies, but it does not fall at high frequencies due to the membrane capacitance and frequency independent terms. The column-mean-square component $S_{R}$ has a higher frequency content than the linear behavior, although less dramatic than the marked scaled resonance of the frequency doubling component $S_{D}$.

In summary, the power spectra of the QSA responses for the $n^{4}$ model are clearly different than those of Markov fluctuations and the power spectra of the linear responses do not generally follow that of the spontaneous fluctuations. Thus, Markov fluctuations are not predicted by small signal evoked linear or quadratic responses. Interestingly, Markov fluctuations at depolarized membrane potentials are dominated by the linear time constant, this is not generally the case for lesser depolarizations, as illustrated between a $5\mathrm{mV}$ and $55\mathrm{mV}$ depolarizations.

\subsection{Observations on the linear fluctuation-dissipation theorem}\label{subsec_fluctuation_dissipation_theorem}

As discussed by \cite{Stevens_1972}, the power $n^{4}$ in Hodgkin-Huxley equations leads to two interpretations of the origin of potassium conductance noise. The first case involves Markov simulations and power spectra with four Lorentzian terms, as illustrated in Fig. \ref{fig:qsa_vc_n4_5mV_M128_noise_stimulated}. The second case involves linearization of the Hodgkin-Huxley equations with a single relaxation time, as illustrated by the term $g_{\mathrm{K}}\left[4n_{0}^{3}\left(V_{0}-V_{\mathrm{K}}\right)\widehat{D_{n}}+n_{0}^{4}\right]$ in Eq. \ref{eq:mauro_hh_admittance}.

The linear fluctuation-dissipation theorem states that the spontaneous fluctuations should have the same kinetic behavior as a response to a small signal, namely a linear response. The fact that Markov simulations have four time constants whereas the linearization has a single relaxation time does not agree with the linear fluctuation-dissipation theorem. Thus, the linear fluctuation-dissipation theorem does not hold for the $n^{4}$ model. This suggests exploring nonlinear extensions of the linear fluctuation-dissipation theorem in the physics literature. However, it is also interesting to consider the case when Markov simulations have fewer time constants, as with the $p_{2}$ model described in this article.

\subsection{Comparison between $n^{4}$ and $p_{2}$ models}\label{subsec_comparison_n4_p2}

\subsubsection{Depolarization of $55\mathrm{mV}$}\label{subsec_depolarization_55}

As discussed above, sequential kinetic models can produce the exponentiation of the $n^{4}$ Hodgkin-Huxley model, so that the probability of the channel opening coincides with $n^{4}$. Similarly, for the $n^{2}$ model, the probability of the channel being open coincides with $n^{2}$. However, the $p_{2}$ model generalizes the rate constants of the $n^{2}$ model, so that the probability of the channel opening is not necessarily an exponentiation. Thus, sequential kinetic models are more general than the Hodgkin-Huxley model, as that they do not require exponentiation. In these models, the nonlinear voltage dependence of neuronal ionic conductances lies essentially in the rate constants between sequential states and not in an exponentiation functional dependence of the conductance gating variable.

If such non-exponentiation-based sequential kinetic models also fit the data, then their nonlinear behavior may more realistically describe the underlying molecular mechanisms of ion channel activity. Indeed, in accordance with the principle of parsimony, these models free themselves from the constraint that the probability of channel opening must be an exponentiation.

\begin{figure}[H]
\centering
\includegraphics[width=14cm]{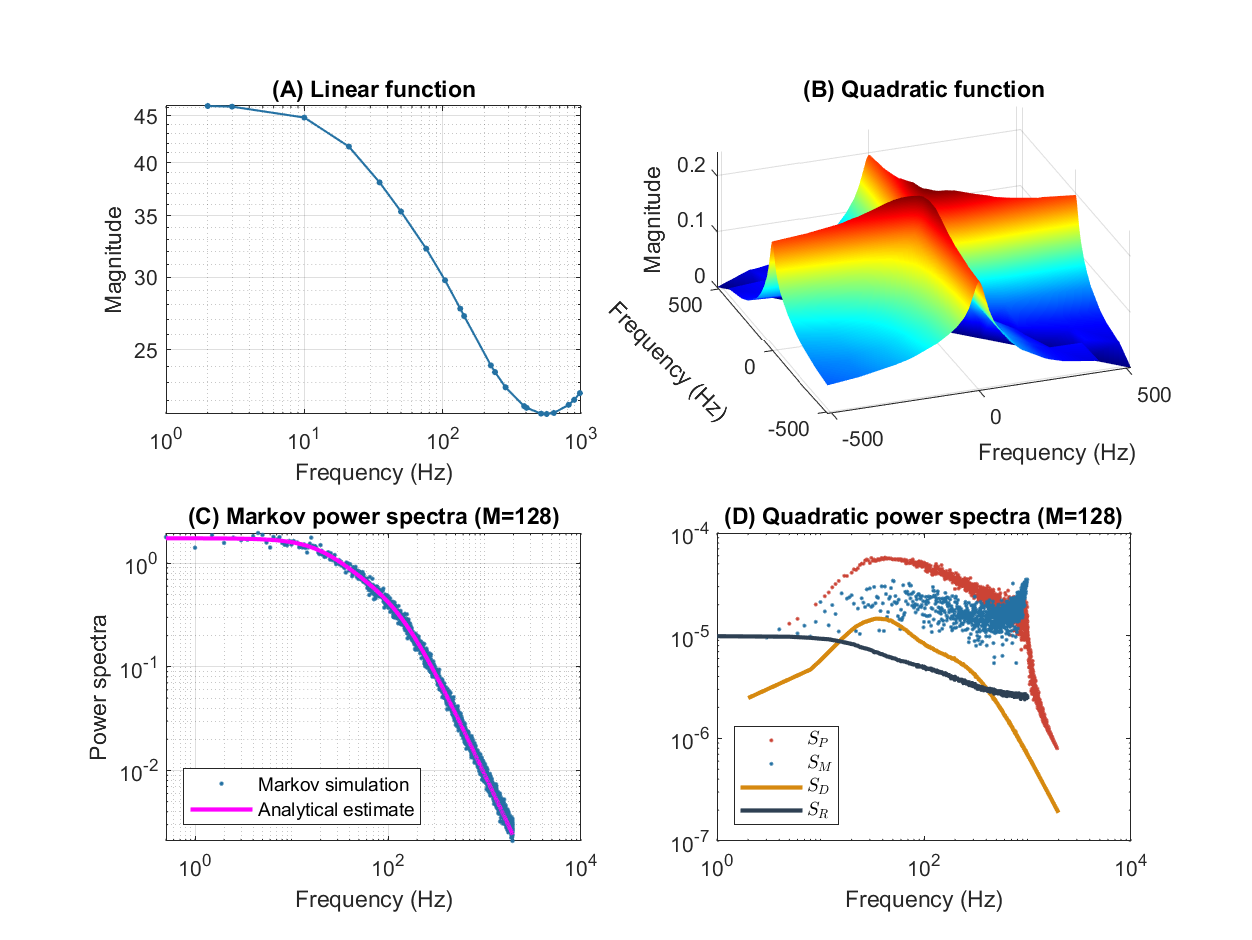}
\caption{\label{fig:qsa_vc_p2_55mV_M128}\textbf{Frequency analysis of the $p_{2}$ model for a $55\mathrm{mV}$ depolarization.} The rate constants have been manually selected to approximate the frequency domain responses of the $n^{4}$ model, which can be compared to Fig. \ref{fig:qsa_vc_n4_55mV_M128}.}
\end{figure}

\begin{figure}[H]
\centering
\includegraphics[width=12cm]{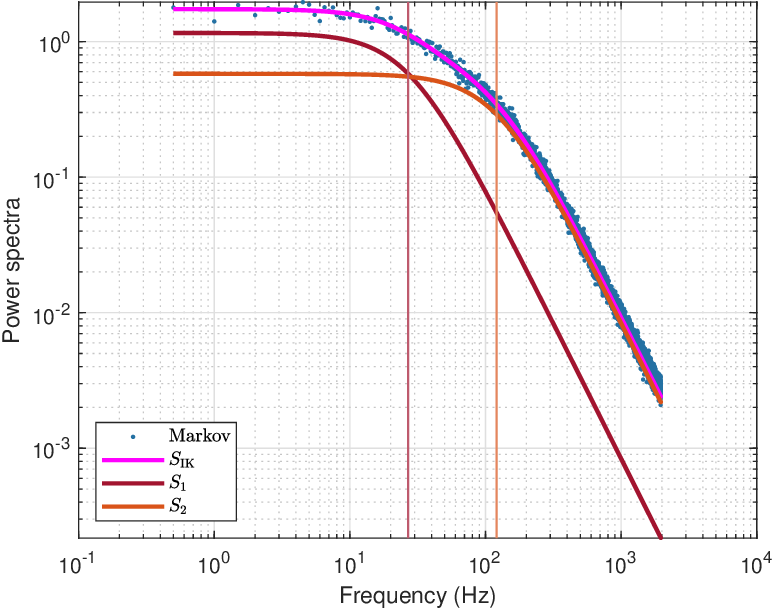}
\caption{\label{fig:qsa_vc_p2_55mV_M128_noise_notstimulated}\textbf{Markov simulation of the $p_{2}$ model for a $55\mathrm{mV}$ depolarization without stimulus.} The rate constants have been manually selected to approximate the frequency domain responses of the $n^{4}$ model, which can be compared to Fig. \ref{fig:qsa_vc_n4_55mV_M128_noise_notstimulated}.}
\end{figure}

Fig. \ref{fig:qsa_vc_p2_55mV_M128} and Fig. \ref{fig:qsa_vc_p2_55mV_M128_noise_notstimulated} show an example of a $p_{2}$ model, in which the rate constants have been manually selected to approximate the frequency domain responses of the $n^{4}$ model, for a depolarization of $55\mathrm{mV}$. As expected, both models show similar resonance behavior for the linear and quadratic responses. However, there are quantitative differences. First, the $S_{D}$ quadratic power spectra of the $p_{2}$ model (Fig. \ref{fig:qsa_vc_p2_55mV_M128}) show not one, but two resonances reflected by a peak and bump, unlike the $n^{4}$ model (Fig. \ref{fig:qsa_vc_n4_55mV_M128}). Second, the Markov fluctuations of the $p_{2}$ model (Fig. \ref{fig:qsa_vc_p2_55mV_M128_noise_notstimulated}) are also different with the indication of more than one time constant illustrated by the inflection just before the final slope of $S_{2}$, unlike the $n^{4}$ model (Fig. \ref{fig:qsa_vc_n4_55mV_M128_noise_notstimulated}) for which $S_{4}$ fits $S_{\mathrm{IK}}$ accurately.

\begin{figure}[H]
\centering
\includegraphics[width=12cm]{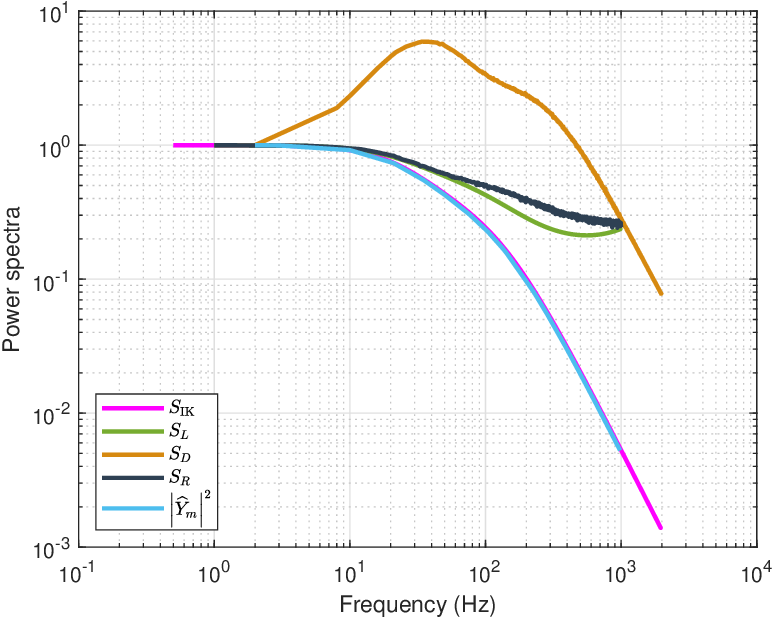}
\caption{\label{fig:qsa_vc_p2_55mV_M128_comparison}\textbf{Power spectral analysis of the $p_{2}$ model for a $55\mathrm{mV}$ depolarization.} The rate constants have been manually selected to approximate the frequency domain responses of the $n^{4}$ model, which can be compared to Fig. \ref{fig:qsa_vc_n4_55mV_M128_comparison}.}
\end{figure}

Fig. \ref{fig:qsa_vc_p2_55mV_M128_comparison} shows the superposition of five curves of the $p_{2}$ model for a $55\mathrm{mV}$ depolarization, using the same presentation as for the $n^{4}$ model in Fig. \ref{fig:qsa_vc_n4_55mV_M128_comparison}. As with the $n^{4}$ model, the modified squared admittance $\left|\widehat{Y}_{m}\right|^{2}$ nearly superimposes on the Markov fluctuations $S_{\mathrm{IK}}$. However, the power spectra are quite different between the two models. The inflection of the Markov fluctuations $S_{\mathrm{IK}}$ is matched by the modified squared admittance $\left|\widehat{Y}_{m}\right|^{2}$, which is consistent with the fact that the analytical expressions for Markov fluctuations and admittance both have two relaxation processes in the $p_{2}$ model. In contrast, the $n^{4}$ model has only one relaxation process due to the linearization of the exponentiation. Similarly, the modified squared admittance $\left|\widehat{Y}_{m}\right|^{2}$ agrees well, at low frequencies, with the linear power spectrum $S_{L}$ and column-mean-square component $S_{R}$. Interestingly, the frequency doubling component $S_{D}$ exhibits a double resonance,suggesting the existence of more than one relaxation process.

\subsubsection{Depolarization of $5\mathrm{mV}$}\label{subsec_depolarization_5}

\begin{figure}[H]
\centering
\includegraphics[width=14cm]{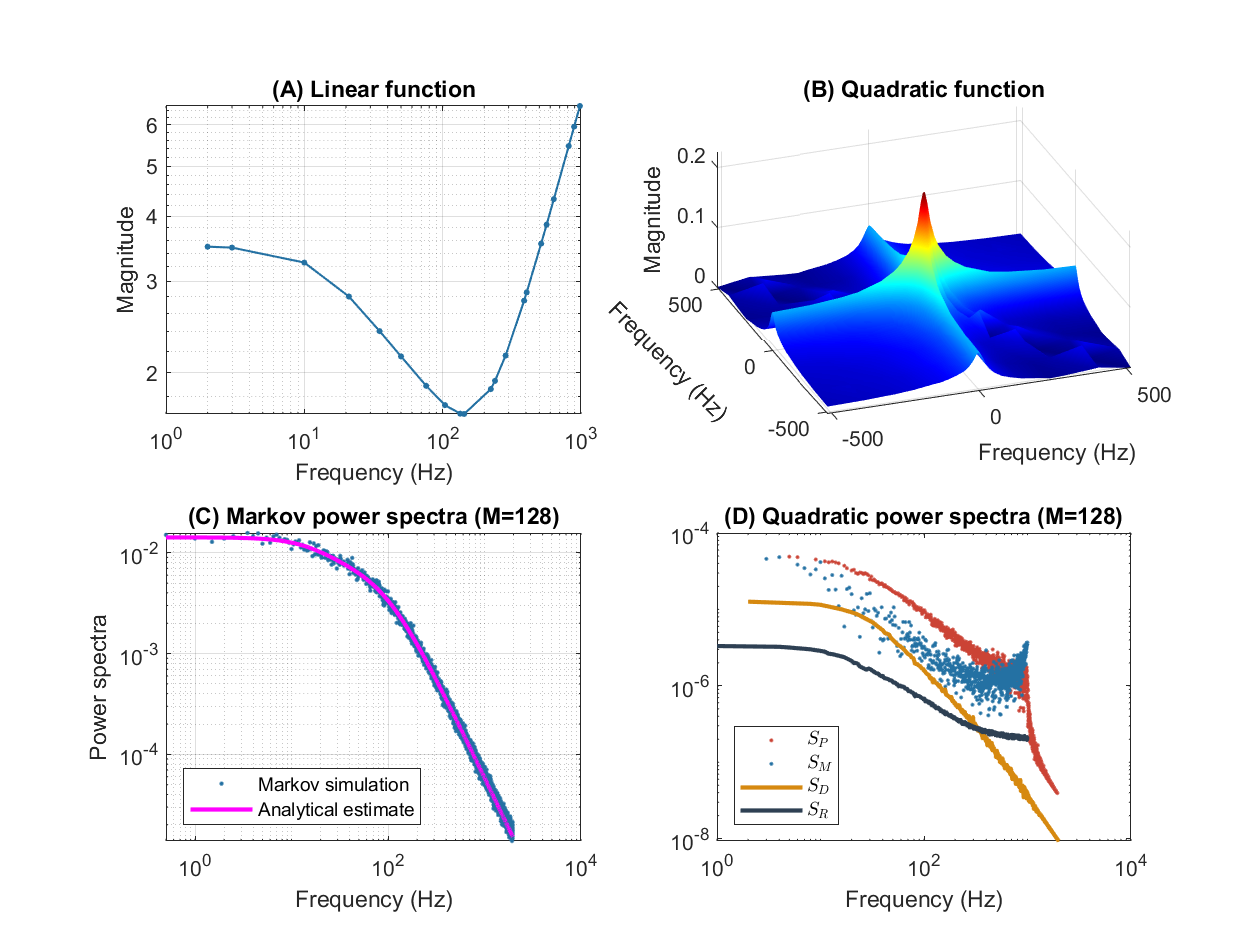}
\caption{\label{fig:qsa_vc_p2_5mV_M128}\textbf{Frequency analysis of the $p_{2}$ model for a $5\mathrm{mV}$ depolarization.} The rate constants have been manually selected to approximate the frequency domain responses of the $n^{4}$ model, which can be compared to Fig. \ref{fig:qsa_vc_n4_5mV_M128}.}
\end{figure}

Fig. \ref{fig:qsa_vc_p2_5mV_M128} shows the same simulations as in Fig. \ref{fig:qsa_vc_p2_55mV_M128} but for a depolarization of $5\mathrm{mV}$. The results are quite similar to the $n^{4}$ model in Fig. \ref{fig:qsa_vc_n4_5mV_M128}. The upper left plot shows a typical low frequency linear admittance anti-resonance. The upper right plot shows the QSA matrix as a 3D representation, which shows no anti-resonance. The lower left plot shows the power spectra of the Markov simulation superimposed on the analytical estimate, both of which reveal the characteristics of a low-pass filter. The lower right plot shows the quadratic power spectra averaged over several ODE simulations, all of these responses exhibit low-pass filter characteristics, but they flatten out by reaching a constant value at high frequencies, with the exception of $S_{D}\left(2\left|\omega_{k}\right|\right)$ at frequency doubling for which the amplitudes decrease at high frequencies.

\begin{figure}[H]
\centering
\includegraphics[width=12cm]{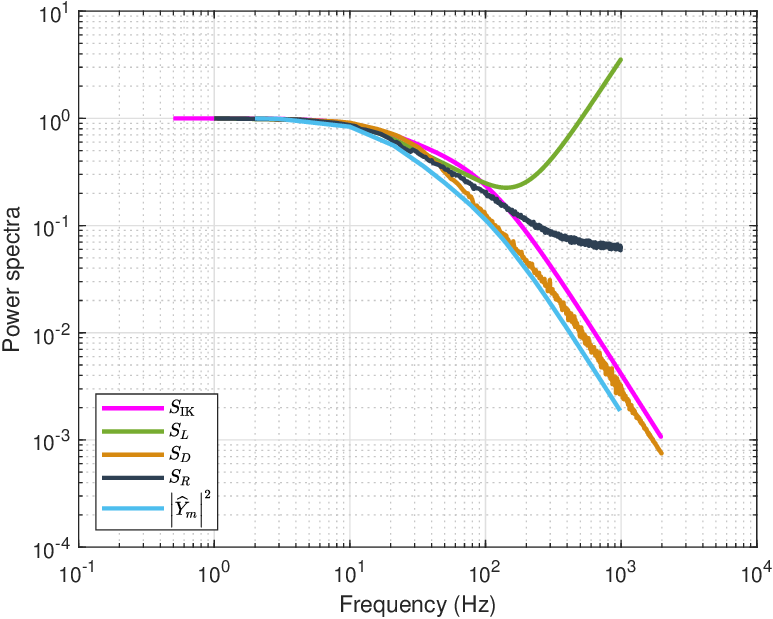}
\caption{\label{fig:qsa_vc_p2_5mV_M128_comparison}\textbf{Power spectral analysis of the $p_{2}$ model for a $5\mathrm{mV}$ depolarization.} The rate constants have been manually selected to approximate the frequency domain responses of the $n^{4}$ model, which can be compared to Fig. \ref{fig:qsa_vc_n4_5mV_M128_comparison}.}
\end{figure}

Fig. \ref{fig:qsa_vc_p2_5mV_M128_comparison} shows that, as with the $n^{4}$ model at $5\mathrm{mV}$ in Fig. \ref{fig:qsa_vc_n4_5mV_M128_comparison}, the linear behavior represented by the modified squared admittance $\left|\widehat{Y}_{m}\right|^{2}$ does not accurately describe quadratic power spectra or Markov fluctuations. In particular, as with the $n^{4}$ model at $5\mathrm{mV}$, the modified squared admittance and quadratic power spectra of the $p_{2}$ model have lower frequency components than the Markov fluctuations.

\section{Discussion}\label{sec_discussion}

Neuronal fluctuations influence both transient and steady state responses, which fundamentally set the thresholds for neuronal impulse propagation. The fluctuating behavior of neurons in the central nervous system can be divided into three categories: (1) subthreshold synaptic bombardment from multiple networks, (2) additional nonlinear components generated for a particular stimulus, and (3) intrinsic spontaneous noise independent of external stimulus. Computational neuroscience is an attempt to understand the network behavior of the nervous system. However, it is highly dependent on neuronal models determined from numerous systems using very restricted data sets. The tools for extracting experimental data are impressive, ranging from current and voltage clamp, single channel and noise analysis, multi-electrode arrays recording activity on a population of cells, imaging techniques to estimate activity in intact systems, transfer function analysis and a variety of nonlinear approaches such as Wiener and Volterra kernels. The QSA method used in this article refines the quantitative characteristics of neuronal models in the frequency domain.

The above simulations shown for the potassium conductance suggest that the non-smooth fluctuations of the QSA power spectra $S_{P}$ and $S_{M}$ generate an alternative kind of noise due to the complexity of nonlinear interactions, which is not identical to the spontaneous fluctuations simulated by a stochastic Markov process. Although each individual QSA characterization uses a limited number of non-overlapping frequencies, this is overcome relatively well by averaging the power spectra for different stimulus sets. The $p_{2}$ sequential model of the ion channel suggests that the fluctuation-dissipation theorem may be, in part, valid since both the linear and fluctuation spectra have similar relaxation times (corner frequencies), which is not the case for the $n^{4}$ model.

QSA analysis of a Markov model requires the averaging of individual trace measurements using the same multi-sinusoidal stimulus, which will converge to the deterministic response if a sufficient number of averages are performed. QSA analysis of individual trace measurements leads to highly variable QSA power spectra, unlike the same procedures performed on deterministic models, which are invariant. As real biological neuronal cells are intrinsically fluctuating, QSA experiments on patch clamped neurons have been performed on averaged trace measurements \citep{Magnani2011}, leading to a deterministic response that averages out the spontaneous fluctuation behavior of ion channels. The above-mentioned QSA power spectra were performed with random QSA stimulus frequencies applied to ODEs in order to compare them with the Markov noise power spectra. Alternatively, a single set of QSA stimulus frequencies can be applied to a set of Markov simulations, in which case the power spectra of the QSA matrix coefficients for the same stimulus can be averaged to determine how prominent the Markov model's responses are at the nonlinear interactive frequencies. This can be called an average QSA Markov noise power spectrum.

\begin{figure}[H]
\centering
\begin{tabular}{ccc}
\includegraphics[width=5cm]{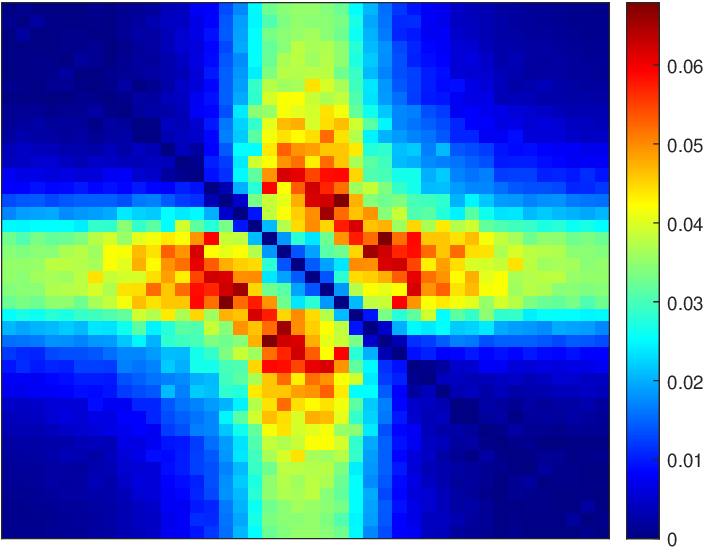} &
\includegraphics[width=5cm]{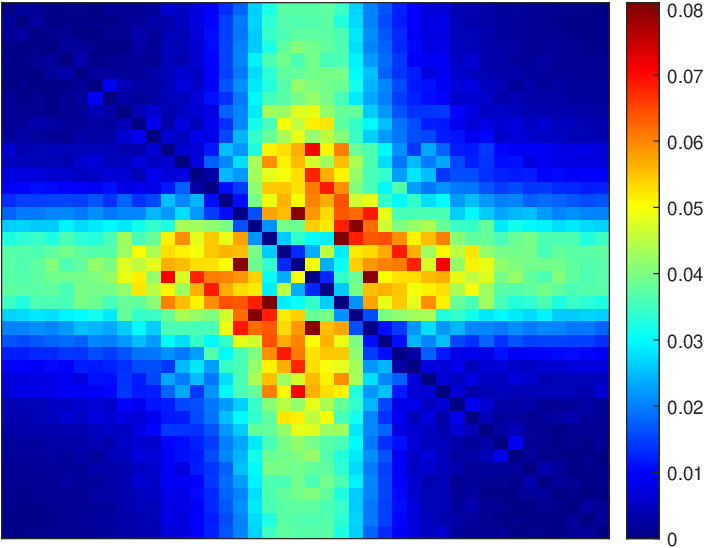} &
\includegraphics[width=5cm]{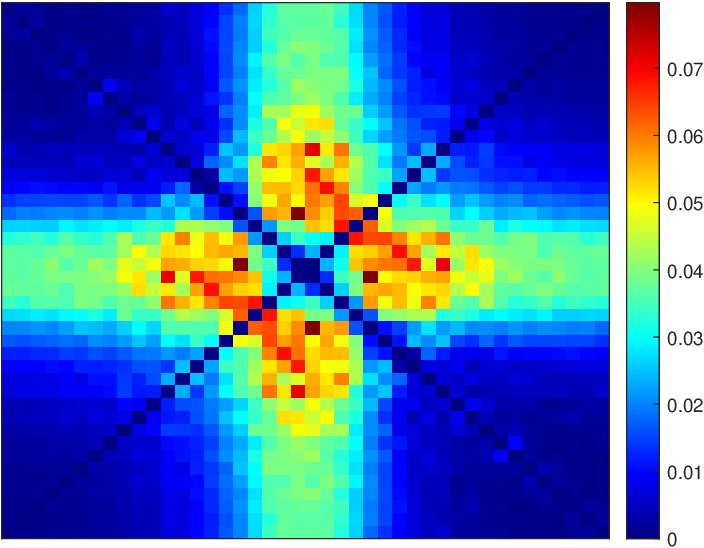} \\
\includegraphics[width=5cm]{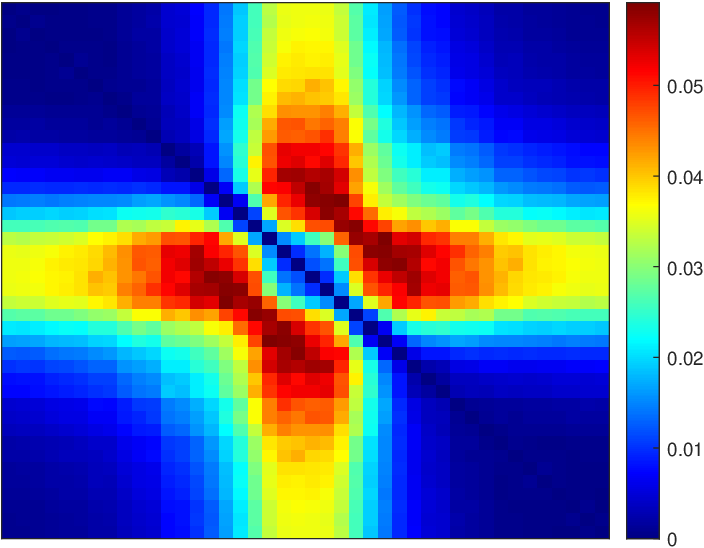} &
\includegraphics[width=5cm]{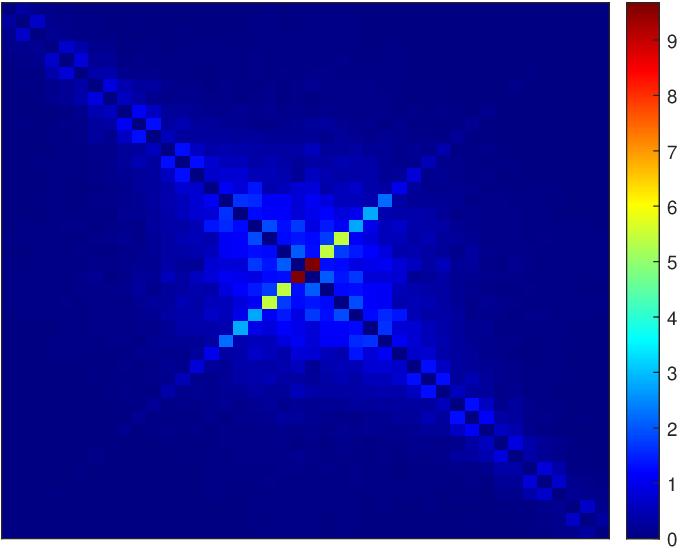} &
\includegraphics[width=5cm]{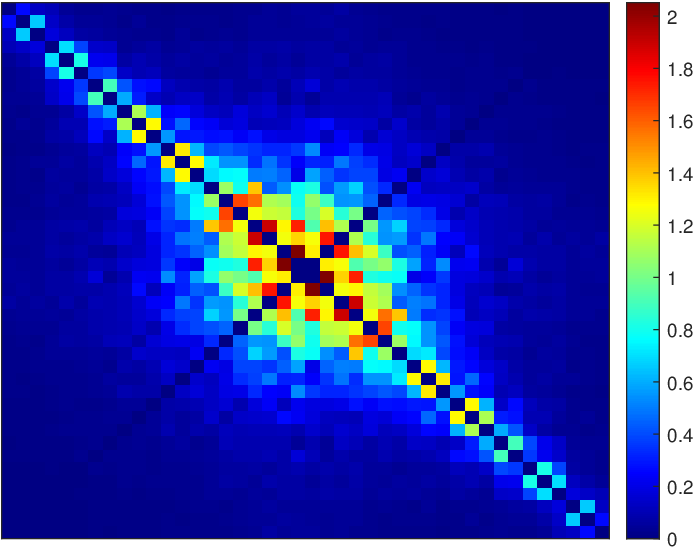} \\
\end{tabular}
\caption{\label{fig:qsa_vc_p2_55mV_4mV_1mV_M16_qsa_noise}\textbf{Power spectra of the QSA matrix coefficients of the $p_{2}$ model for a $55\mathrm{mV}$ depolarization.} \textbf{(Upper row)} Stimulus amplitude $4\mathrm{mV}$. \textbf{(Lower row)} Stimulus amplitude $1\mathrm{mV}$. \textbf{(Left column)} Deterministic ODE equations. \textbf{(Middle column)} QSA analysis performed on each individual Markov simulation for $16$ iterations. \textbf{(Right column)} Identical to the middle column but with the diagonal (harmonic frequencies) removed (set to zero).}
\end{figure}

Fig. \ref{fig:qsa_vc_p2_55mV_4mV_1mV_M16_qsa_noise} shows the power spectra of the QSA matrix coefficients of the $p_{2}$ model for a $55\mathrm{mV}$ depolarization, comparing two different stimulus amplitudes $4\mathrm{mV}$ (top) and $1\mathrm{mV}$ (bottom). Stimulus amplitudes are relatively larger than those used previously in deterministic analyses, as quadratic responses must overcome spontaneous fluctuations, otherwise the quality of noisy signals is too degraded. The two plots on the left column represent the power spectra of the usual deterministic ODE equations, which serve as a reference. As expected, they are similar at $4\mathrm{mV}$ and $1\mathrm{mV}$. The two plots in the middle column represent the power spectra of the QSA Markov noise (QSA analysis performed on each individual Markov simulation). Surprisingly, the result is similar to that of ODE at $4\mathrm{mV}$, but dramatically different at $1\mathrm{mV}$. In particular the enhanced diagonal follows harmonic frequencies. The right column shows the same power spectra as the middle column, but with the diagonal (harmonic frequencies) removed (set to zero). The result remains dramatically different from the ODE at $1\mathrm{mV}$, which means that both the diagonal and cross-terms responses represent another quadratic function. Namely, spontaneous Markov fluctuations modify the neuronal quadratic function as the stimulus amplitude decreases. It is interesting to note that harmonic frequencies of the QSA Markov noise are also enhanced at $4\mathrm{mV}$, but the matrix is essentially similar to that of ODE. The enhanced diagonal may be due to the fact that the modified squared admittance $\left|\widehat{Y}_{m}\right|^{2}$ predicts the Markov fluctuations $S_{\mathrm{IK}}$ at a $55\mathrm{mV}$ depolarization, as shown in Fig. \ref{fig:qsa_vc_p2_55mV_M128_comparison}. Each squared frequency component generates frequency doubling. Thus, Markov fluctuations contribute to the harmonic frequencies in the QSA analysis of each individual Markov simulation, as well as at, to a lesser extent, to the interactive frequencies since the Markov noise cloud is not a smooth curve.

\begin{figure}[H]
\centering
\begin{tabular}{ccc}
\includegraphics[width=5cm]{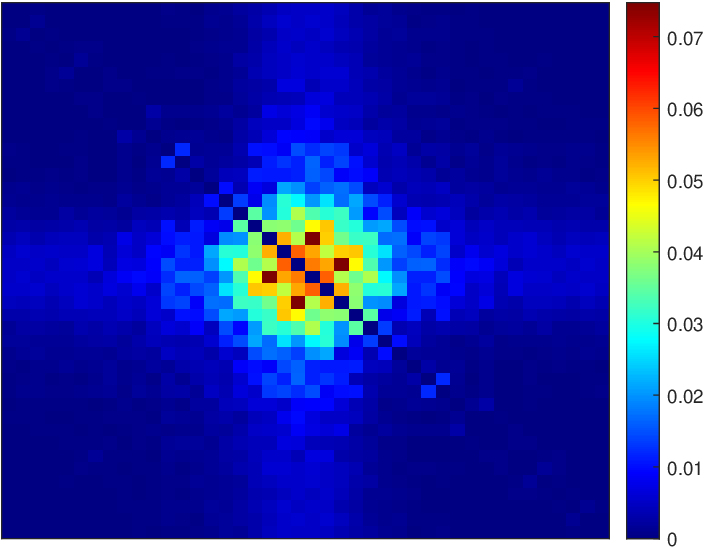} &
\includegraphics[width=5cm]{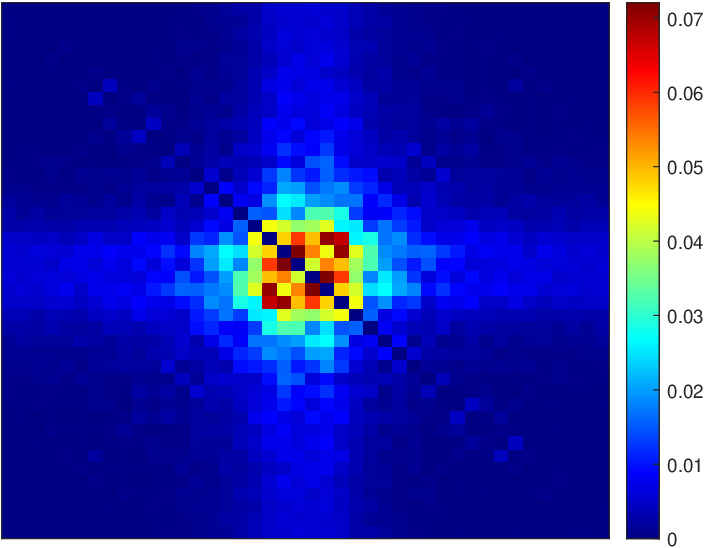} &
\includegraphics[width=5cm]{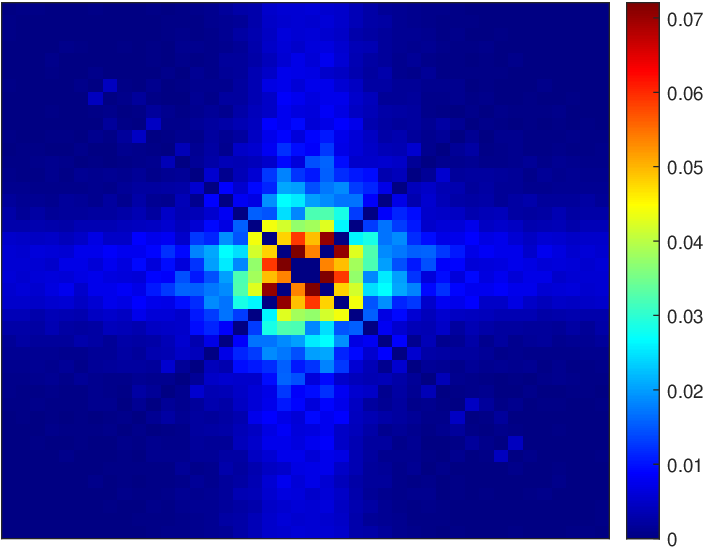} \\
\includegraphics[width=5cm]{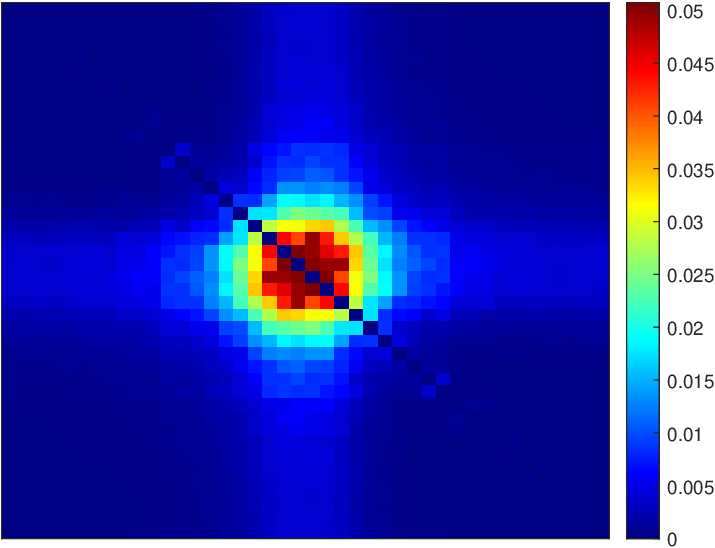} &
\includegraphics[width=5cm]{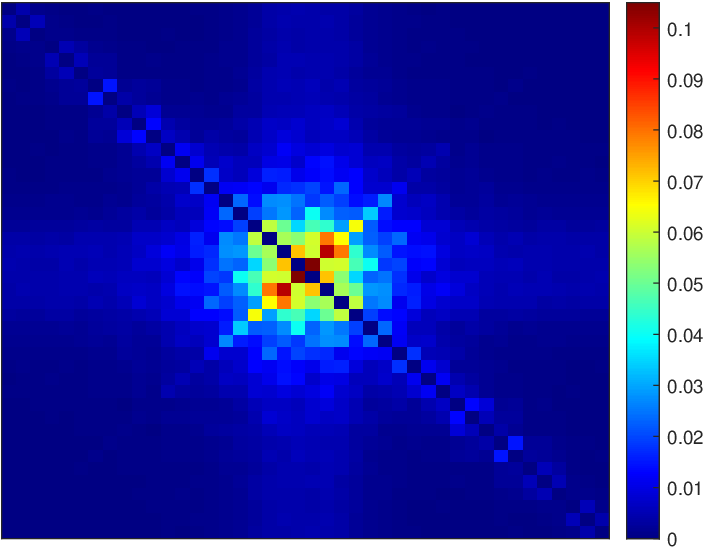} &
\includegraphics[width=5cm]{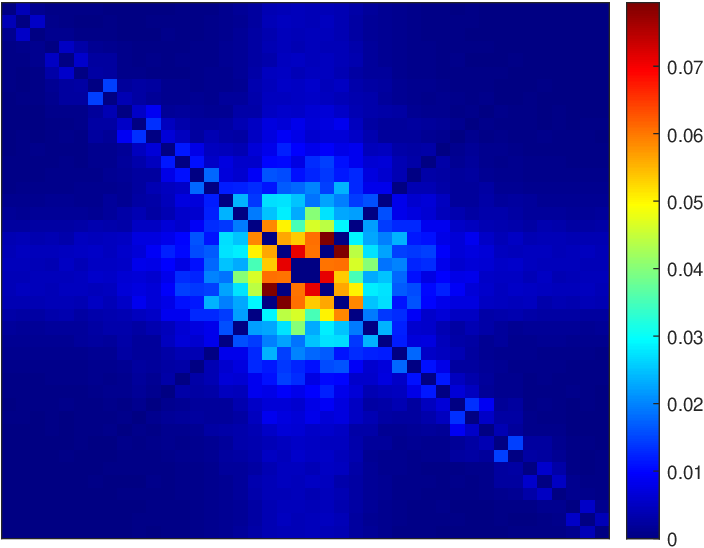} \\
\end{tabular}
\caption{\label{fig:qsa_vc_p2_5mV_4mV_1mV_M16_qsa_noise}\textbf{Power spectra of the QSA matrix coefficients of the $p_{2}$ model for a $5\mathrm{mV}$ depolarization.} \textbf{(Upper row)} Stimulus amplitude $4\mathrm{mV}$. \textbf{(Lower row)} Stimulus amplitude $1\mathrm{mV}$. \textbf{(Left column)} Deterministic ODE equations. \textbf{(Middle column)} QSA analysis performed on each individual Markov simulation for $16$ iterations. \textbf{(Right column)} Identical to the middle column but with the diagonal (harmonic frequencies) removed (set to zero).}
\end{figure}

Fig. \ref{fig:qsa_vc_p2_5mV_4mV_1mV_M16_qsa_noise} supports this hypothesis as illustrated with the power spectra of the QSA matrix coefficients of the $p_{2}$ model for a $5\mathrm{mV}$ depolarization, comparing two different stimulus amplitudes $4\mathrm{mV}$ (top) and $1\mathrm{mV}$ (bottom). At this low depolarization level, the modified squared admittance $\left|\widehat{Y}_{m}\right|^{2}$ no longer predicts the Markov fluctuations $S_{\mathrm{IK}}$, as shown in Fig. \ref{fig:qsa_vc_p2_5mV_M128_comparison}. In this case, the diagonal of the QSA Markov noise is smaller and the power spectra of the QSA matrix coefficients are similar to those of the ODE at $4\mathrm{mV}$ and $1\mathrm{mV}$. Nevertheless, the diagonal remains somewhat enhanced in the QSA Markov noise, likely because Markov fluctuations and modified squared admittance $\left|\widehat{Y}_{m}\right|^{2}$ have a similar appearance.

\begin{figure}[H]
\centering
\includegraphics[width=12cm]{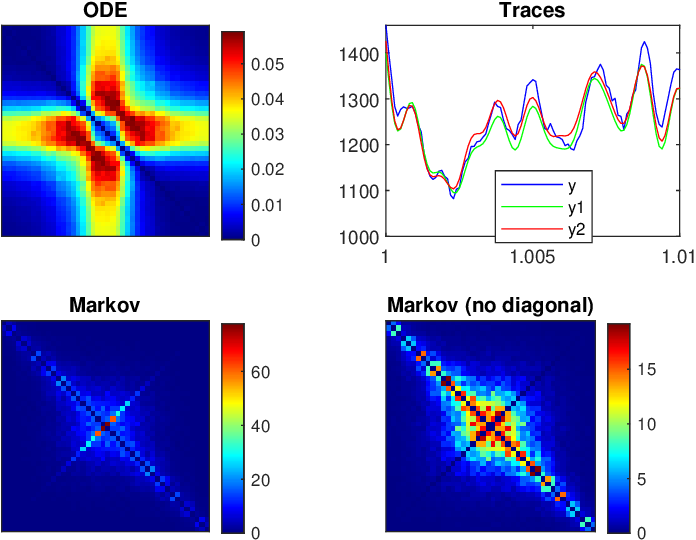}
\caption{\label{fig:qsa_vc_p2_55mV_1mV_Area_M16_qsa_noise_A50}\textbf{Power spectra of the QSA matrix coefficients of the $p_{2}$ model for a membrane surface area $A_{\mathrm{K}}=50\mathrm{\mu m}^{2}$, a $55\mathrm{mV}$ depolarization and a stimulus amplitude $1\mathrm{mV}$.} \textbf{(Top left)} Deterministic ODE equations. \textbf{(Top right)} A single trace (blue curve) from Markov simulations in time domain ($\mathrm{s}, \mathrm{pA}$) with linear (green curve) and quadratic (red curve) analyses. \textbf{(Bottom left)} QSA analysis performed on each individual Markov simulation for $16$ iterations. \textbf{(Bottom right)} Identical to the bottom left but with the diagonal (harmonic frequencies) removed (set to zero).}
\end{figure}

\begin{figure}[H]
\centering
\includegraphics[width=12cm]{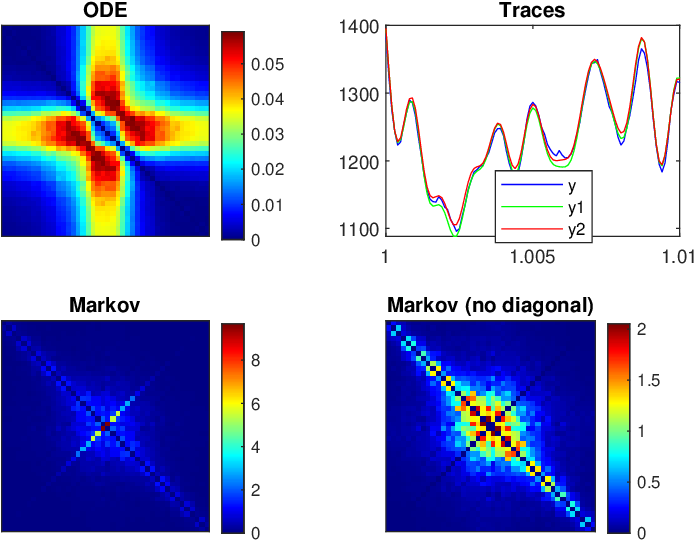}
\caption{\label{fig:qsa_vc_p2_55mV_1mV_Area_M16_qsa_noise_A500}\textbf{Power spectra of the QSA matrix coefficients of the $p_{2}$ model for a membrane surface area $A_{\mathrm{K}}=500\mathrm{\mu m}^{2}$, a $55\mathrm{mV}$ depolarization and a stimulus amplitude $1\mathrm{mV}$.} The plots were generated using the same presentation as in Fig. \ref{fig:qsa_vc_p2_55mV_1mV_Area_M16_qsa_noise_A50}.}
\end{figure}

\begin{figure}[H]
\centering
\includegraphics[width=12cm]{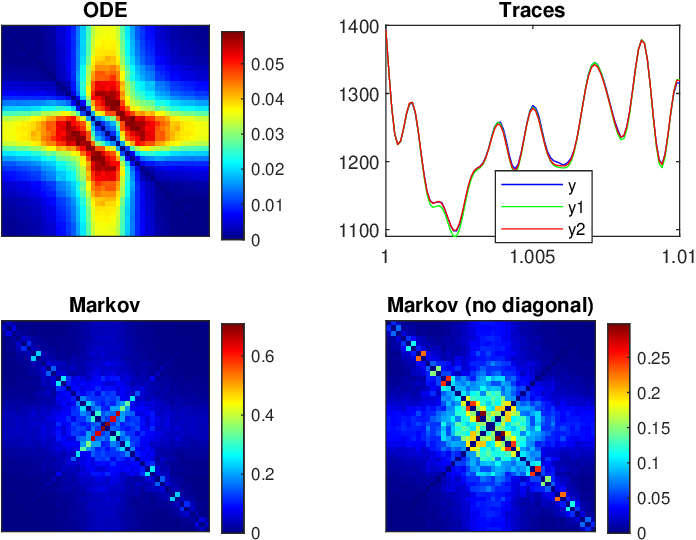}
\caption{\label{fig:qsa_vc_p2_55mV_1mV_Area_M16_qsa_noise_A5000}\textbf{Power spectra of the QSA matrix coefficients of the $p_{2}$ model for a membrane surface area $A_{\mathrm{K}}=5000\mathrm{\mu m}^{2}$, a $55\mathrm{mV}$ depolarization and a stimulus amplitude $1\mathrm{mV}$.} The plots were generated using the same presentation as in Fig. \ref{fig:qsa_vc_p2_55mV_1mV_Area_M16_qsa_noise_A50}.}
\end{figure}

\begin{figure}[H]
\centering
\includegraphics[width=12cm]{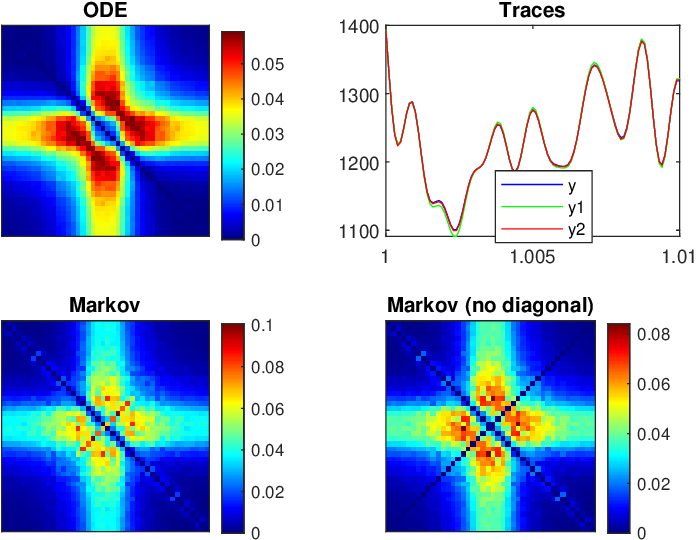}
\caption{\label{fig:qsa_vc_p2_55mV_1mV_Area_M16_qsa_noise_A50000}\textbf{Power spectra of the QSA matrix coefficients of the $p_{2}$ model for a membrane surface area $A_{\mathrm{K}}=50000\mathrm{\mu m}^{2}$, a $55\mathrm{mV}$ depolarization and a stimulus amplitude $1\mathrm{mV}$.} The plots were generated using the same presentation as in Fig. \ref{fig:qsa_vc_p2_55mV_1mV_Area_M16_qsa_noise_A50}.}
\end{figure}

Fig. \ref{fig:qsa_vc_p2_55mV_1mV_Area_M16_qsa_noise_A50}, \ref{fig:qsa_vc_p2_55mV_1mV_Area_M16_qsa_noise_A500}, \ref{fig:qsa_vc_p2_55mV_1mV_Area_M16_qsa_noise_A5000}, \ref{fig:qsa_vc_p2_55mV_1mV_Area_M16_qsa_noise_A50000} reinforce this interpretation by varying the surface area of the membrane of the $p_{2}$ model for a $55\mathrm{mV}$ depolarization and stimulus amplitude $1\mathrm{mV}$. The larger the surface area of the membrane, the lower the noise effect. The smallest areas $A_{\mathrm{K}}=50\mathrm{\mu m}^{2}$, $A_{\mathrm{K}}=500\mathrm{\mu m}^{2}$, $A_{\mathrm{K}}=5000\mathrm{\mu m}^{2}$ show a behavior similar to Fig. \ref{fig:qsa_vc_p2_55mV_4mV_1mV_M16_qsa_noise}, that is to say spontaneous Markov fluctuations modify the neuronal quadratic function. However, the largest area $A_{\mathrm{K}}=50000\mathrm{\mu m}^{2}$ shows a behavior that partially recovers the ODE and better without the diagonal. This suggests that increasing the stimulus amplitude or decreasing the noise amplitude have similar effects on the quadratic function expressed. However, the two approaches are not equivalent since the amplitude of the stimulus tends to be modulated by the inputs of a neuron while the amplitude of the noise depends on the anatomy of a neuron.

This suggests that responses to neuronal stimuli would be different for quiet versus noisy neurons. Quiet neurons would exhibit complex nonlinear frequency responses, whereas the nonlinear responses of noisy neurons would tend to have additional frequency components, particularly at harmonic frequencies of the stimulus. Thus, individual neurons within a neuronal network would process input stimuli according to background synaptic activity, which is likely to be the main determinant of neuronal noise.

Although the Markov model is considered the gold standard for simulating neuronal fluctuations, other approaches can also be used. This is a complex subject as indicated by \cite{ODonnell2014} who have shown that a variety of stochastic considerations for the Hodgkin-Huxley model itself show significant, but sometimes subtle, differences. In particular, It\^{o} processes are defined by stochastic differential equations (SDEs) that can depend on past states, whereas Markov processes depend only on the present state. Thus, It\^{o} processes can be more general than Markov processes. As a particular case, It\^{o} diffusion is similar to the Langevin equation and satisfies the Markov property. Such SDE models are discussed, for instance, by \cite{Goldwyn2011a,Goldwyn2011b}.

The $p_{2}$ model presented in this article is an extension of the Hodgkin-Huxley model, which is based more on the chemical kinetic approach. Unfortunately, experimental measurements of neuronal noise have not fully resolved the question of which type of model best fits measured data.

The validity of a model with exponentiation (such as $n^{4}$) versus a sequential chemical state model (such as $p_{2}$) as a correct description of ion channel kinetics needs to be determined experimentally. If power spectra are fundamentally different, from a kinetic point of view, from the small signal linear responses, then this behavior would argue in favor of a deterministic model with exponentiation (such as $n^{4}$) rather than a Markov model (such as $p_{2}$). The experimental literature shows clear differences between fluctuation power spectra and linear responses.

Squid axon potassium conductance responses to increasing step amplitudes show activation delays described by power functions with increasing exponents as large as $n^{25}$. Decreasing the step amplitudes abolishes the delay as one would expect for a power function. However, this requires a variable exponent that depends on the previous history, which seems rather empirical compared to more realistic models of sequential chemical states, typically used for chemical reactions.

In general, measured fluctuation power spectra (from squid axons and nodes of Ranvier) have been empirically described by Markov power spectra based on the Hodgkin-Huxley model with an exponentiation formalism and, furthermore, direct comparison of fluctuation power spectra for sodium conductance with linear impedance measurements do not agree \citep{Fishman_1983}.

Thus, it is quite clear that the linear behavior of real axons does not predict the measured noise power spectra, and that Markov type fluctuations appear to be consistent with data from a number of different neuronal preparations.

More recently, \cite{Andreozzi2019} have carried out a different type of comparison between the Hodgkin-Huxley and kinetic formalisms with reference to experimental measurements on sodium ion channels.

In conclusion, neurons clearly have linear and nonlinear responses to input stimuli. Linear responses are not just mirror images of the stimulus, they are clearly capable of enhancing certain frequencies due to linear resonance behavior. The linear response of resonance for the Hodgkin-Huxley potassium conductance is only present if the steady state value of $n$ is voltage dependent, namely $\frac{dn_{\infty}}{dV_{0}}>0$, i.e. the resonance is a linear response that depends on the nonlinear property of the conductance. In general, nonlinear responses are considerably more complex, as shown by the presence of new interactive frequencies in the neuronal response that are not present in the input signal. The fluctuations present in neurons are due to voltage-dependent random mechanisms for which probabilities are controlled both linearly and nonlinearly and, as suggested above, ongoing synaptic activity can alter the nature of nonlinear responses (in the sense that QSA matrix can reflect fluctuations for small stimulus amplitudes).

\bibliography{sn-bibliography}

\end{document}